# Circularly-Symmetric Alternating Optical Vortex Lattices and their Focusing Characteristics

Dadong Liu and Li-Gang Wang*

*School of Physics, Zhejiang University, Hangzhou, 310058, China*
*Corresponding author
E-mail address*: lgwang@zju.edu.cn (L.-G. Wang).

**ABSTRACT**

Generating controllable optical vortex (OV) lattices (OVLs) with arbitrary-order topological charge (TC) and superior optical characteristics are highly desirable for various applications. Here, we report an experimental realization of circularly- symmetric OVLs with alternating positive and negative TCs of order $\pm n$, referred to the $n$th-order circularly-symmetric alternating OVL (CSAOVL). The focusing fields of such CSAOVLs exhibit the very interesting patterns with a period of 4 for different $n$ values, and their intensity and phase distributions can be regulated using the radial and azimuthal parameters. Particularly, the formed central bright spot in the focusing fields of those CSAOVLs with $n$ equaling to a multiple of 4 is observed to be much smaller than a Gaussian spot. The results can promote significantly the exploration of structured OVLs and provide potential applications in particle manipulation, imaging, and the realization of complex interactions between optical lattices and microscope particles such as atoms and micro- or nano-particles.

## 1. Introduction

Optical vortices (OVs) [1-3], carrying distinctive attributes such as orbital angular momenta, phase singularities and hollow-intensity structures [4], bring a diverse array of novel applications encompassing optical tweezers [5,6], optical communications [7], quantum physics [8,9], optical imaging [10,11], and optical processing [12]. Besides individual OV, OV lattices (OVLs) consisting of a network of optical vortices are further gaining increasing attention owing to their rich physical properties and offer enhanced advantages in many fields. For instance, OVLs exhibit superior capabilities for manipulating multiple particles within the realm of optical traps [13,14]. Their utility also extends to diverse domains, like multi-channel optical communications [15], micro-optomechanical pumps [16], quantum processing [17], and ultracold atoms trapping [18,19].

Several methodologies are proposed to generate OVLs. Notably, diffractive optical elements, such as Dammann gratings [20,21], spatial light modulators (SLMs) [22-25], and metasurfaces [26-28], are prevalent technologies in this regard. Advanced approaches involve micro-structure materials, utilizing topological defects in nematic liquid crystals [29] and self-assembled structures in smectic, nematic, and cholesteric liquid crystals [30-32]. OVLs are also generated via multi-beam interference, encompassing the interference of plane or spherical waves [33,34], Laguerre–Gaussian beams [18], Ince-Gaussian beams [35], Bessel beams [36], and perfect OVs [37]. To date, various structures of OVLs have been developed, including triangle [38], square [39-41], hexagonal [42], "flower" [43], "bear" [44], and arbitrary curvilinear OVLs [25,45].

The aforementioned investigations of OVLs have predominantly concentrated on their generation methodologies [18,20-37], diverse structural manifestations [25,38-45], and practical applications [13-19]. Researchers have also delved into the propagation properties of OVLs mainly consisting of alternating $\pm 1$ topological charges (TCs), which result in the formation of ordered arrays of bright beams with flat phase profiles at the focal plane [39,42]. However, the study on OVLs incorporating high-order OVs remains relatively little attention due to the challenge of their complex dynamics. In our recent work [41], we reported the tunable square OVLs with alternating arbitrary-order $\pm n$ TCs, and demonstrated their flexible capabilities for manipulating finite square optical arrays with intriguing patterns or optical defects. It is very interesting to disclose the role of the TC value in square OVLs as a new freedom to control the focusing patterns of optical arrays [41]. Here, we propose a novel class of arbitrary order OVL with circularly-symmetric alternating $\pm n$ TCs, which can be called as the $n$th-order circularly-symmetric alternating OVL (CSAOVL). To the best of our knowledge, such lattices have not been reported previously in literature. We experimentally generate these CSAOVLs and demonstrate their focal properties, forming the circularly-symmetric intensity patterns with a period of 4 as the order $n$ increases. Notably, under the particular values when $n$ is equal to a multiple of 4, the focal field always becomes a bright smaller spot in comparison to a focusing Gaussian beam (GB). Our experimental results show an optimal ~26% reduction in the central spot compared with a GB. All experimental results are in good agreement with theoretical predictions.

## 2. Arbitrary-order CSAOVL and its experimental generation

In order to consider circularly-symmetric OVLs, the initial light field of the $n$th-order CSAOVL is written as follows

$$E_i^{(n)}(r,\varphi) = G(r)[\cos(ar) + i\sin(b\varphi)]^n, \tag{1}$$

where $r$ and $\varphi$ are the radial and azimuthal directions of the cylindrical coordinates, $G(r)$ is the amplitude of an incident host beam, $n = 1,2,3,\cdots$ is a positive integer representing the absolute value of alternated positive and negative TCs due to the sign changes of $\cos(ar)$ and $\sin(b\varphi)$. For simplicity, $a$, $b$ are assumed to be positive real numbers that control the amplitude and phase modulations along the $r$ and $\varphi$ directions. The OVLs in Eq. (1) exhibit a series of concentrically-circular OV distributions of radii $(2p-1)\pi/(2a)$, where $p = 1, 2, 3, \cdots$ is the positive integer that indicates the $p$th circle. The difference of radii between two adjacent circles is $\pi/a$. The parameter $b$ controls the azimuthal modulation. Here, we only consider the cases of $b$ being a positive integer, thus on each circle there are $2b$ vortices including $b$ positive vortices and $b$ negative vortices, which are alternately and evenly distributed on these circles. The initial field has a rotational symmetry with amplitude and phase distributions possessing $2b$- and $b$-fold rotational symmetries, respectively. The amplitude and phase modulations of the power function in Eq. (1) for $n = 1,2,3,4$ are illustrated in Fig. 1(a). Clearly, as $n$ increases, the structure of vortex locations remains unchanged, but the degree of amplitude/phase modulations is strengthened since the TC value of each OV in lattices increases.

In experiment, the host beam on an SLM is usually a fundamental GB. Without loss of generality, we consider $G(r) = \exp(-\dfrac{r^2}{w_0^2})$ with $w_0$ the initial beam width of the incident GB. Figure 1(b) plots the measured intensity distributions of these generated CSAOVLs at the generation plane ($z$=0), and the intensity profiles exhibit regular circularly-symmetric structures. For $n = 1$, the TCs of these alternated OVs are ±1, resulting in small dark regions among bright spots. With an increasing $n$, the dark areas among bright spots become larger. We will see later that as the modulation is enhanced by increasing $n$, the evolutions of CSAOVLs show interesting properties during propagation. The evolution of such CSAOVLs in linear optical systems can be predicted by the Collins formula [46,47]. The detailed formula can be found in Refs. [41,48] or in Section A of the Supplemental Materials.

To verify the propagation and focusing properties of CSAOVLs, we experimentally generate CSAOVLs utilizing a phase-only SLM (Holoeye, PLUTO-2-NIR-015). The SLM is illuminated by the GB emitting from a linearly-polarized He-Ne laser with wavelength 632.8 nm. Figure 1(c) shows the experimental setup for both generating CSAOVLs and measuring their intensity evolutions. The beam is expanded with the width $w_0 \approx 1.63$ mm. Following the complex-amplitude modulation approach in Ref. [49], the SLM loads computer-generated holograms based on the angular spectrum of Eq. (1). The first-order diffraction beam is selected using an aperture that is strategically placed at the back focal plane of a 2-$f$ lens ($L_0$) system, which is also the generation (or initial) plane of CSAOVLs (*i.e.*, $z = 0$). Another lens $L$ with its focal length $f = 500$ mm is positioned at a distance equal to $f$ from the aperture. Through the latter lens system, the propagation and focusing effects of arbitrary-order CSAOVL can be observed. The resulting intensities are recorded by a movable 12-bit CMOS camera, through changing the distance $z$ between the initial plane and the camera. In this work, we take $w_0 = 1.63$ mm, and $f = 500$ mm in both experiments and theories.

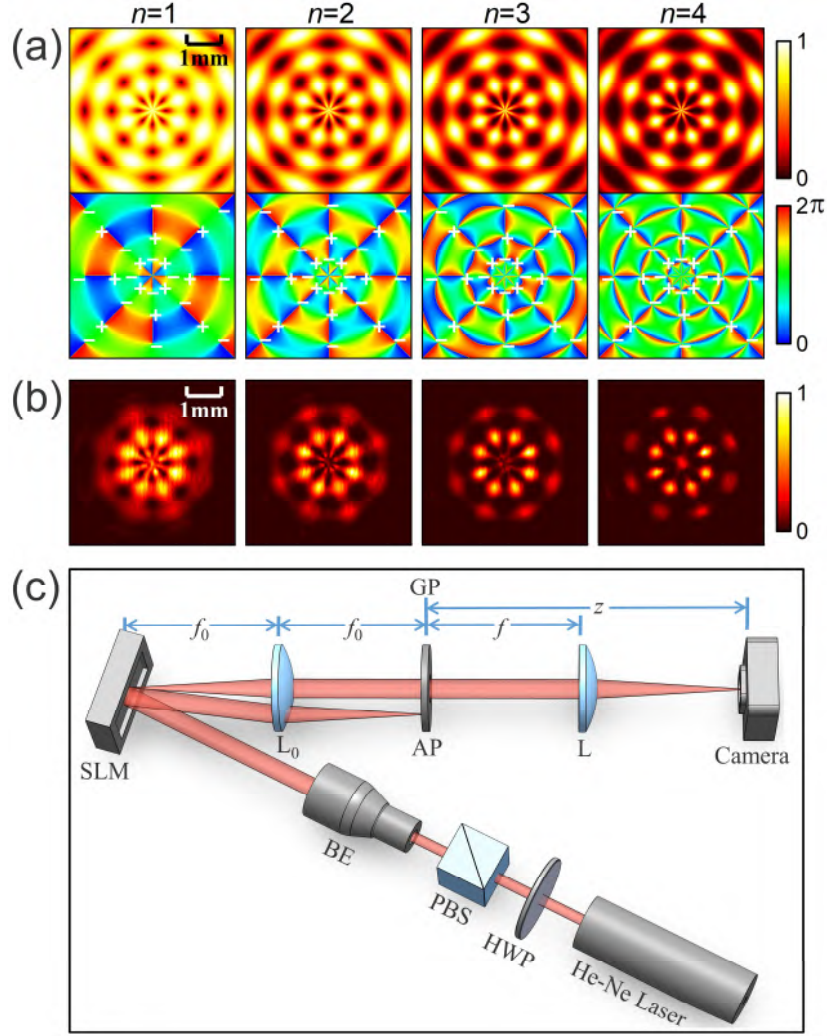

**Fig. 1.** Arbitrary-order CSAOVL and experimental setup. (a) Amplitude (upper) and phase (lower) modulations of the $n$th-order CSAOVL with $n$=1, 2, 3, and 4. Here, the positive and negative vortices are signed by "+" and "-", respectively. (b) The corresponding experimental intensity distributions of the CSAOVLs at the generation plane. In (a-b), the parameters are $a$ = 4 mm$^{-1}$ and $b$ = 4. (c) Schematic of experimental setup. The generation plane (GP) of CSAOVLs is located at the back focal plane of the former 2-$f$ lens (L$_0$) system and the output intensity is measured at the distance $z$ from the GP in the latter lens system (L). Other notations include: HWP, half-wave plate; PBS, polarized beam splitter; BE, beam expander; SLM, spatial light modulator; L$_0$, the lens with $f_0$ = 500mm; L, the lens with $f$ = 500mm; AP, aperture.

## 3. Results and discussions

Figure 2 illustrates the experimental intensity evolutions for different CSAOVLs at various propagation distances ($z$) in a 2-$f$ lens system. Clearly, as $z$ approaches the focal plane ($z$=1000 mm), the intensity patterns initially contract, and then the bright spots exhibit an apparent merging tendency. These evolution characteristics are in good agreement with theoretical calculations, and these intricating patterns arise from the competition and merging mechanism among positive and negative OVs in CSAOVLs. More experimental and theoretical results can be found in Sections B and C of the Supplemental Materials.

For $n$ = 1, 2 and 3, near the focal plane, the central bright spot in these focusing CSAOVLs gradually diminishes, and the focal patterns form specific circular optical arrays with a dark core at the center. Interestingly, for $n$ = 4, the energy of light gradually condenses into the central spot that preserves up to the focal plane, meanwhile the peripheral bright spots become dim as approaching to the focal plane. In Fig. 2(b), the corresponding phase distributions at the focal plane are measured by using the interference method [50]. Via a careful identification, we confirm that the phase singularities of vortex cores or the fast-changing π-phase-like edge dislocations in phase distributions induce dark-field structures at the focal plane. The distributions of these bright spots are resulted from the transverse energy flows of light near the focal plane, see Fig. 2(c). Only when $n$=4 or $4m$ with $m$ = 1, 2, 3, ⋯ being positive integers, the transverse energy flow continues towards the center and forms a bright spot at the focal plane. More information about the phase structures and transverse energy flows can be seen in Section B of the Supplemental Materials.

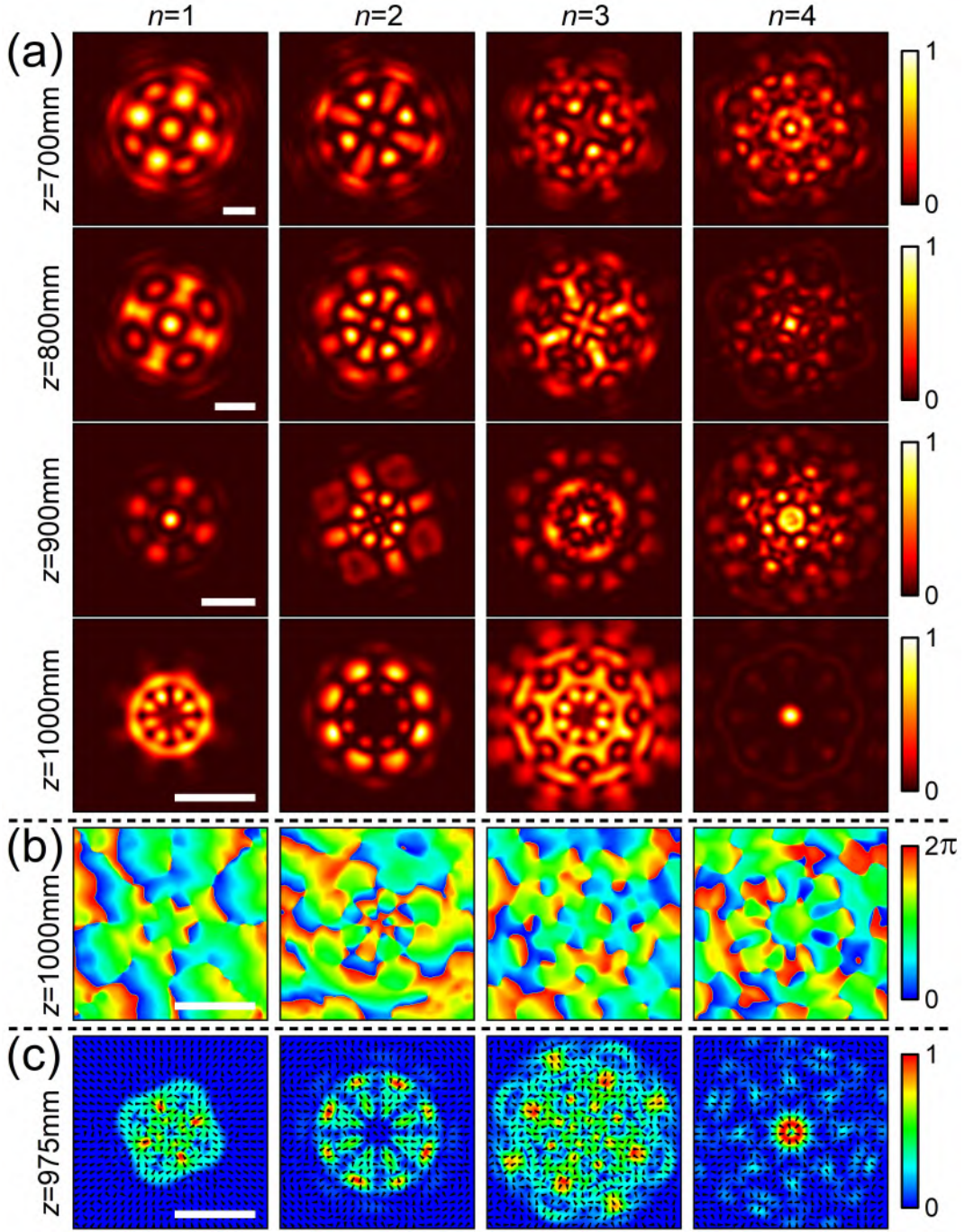

**Fig. 2.** Evolutions of four lowest-order CSAOVLs in a 2-$f$ lens system. (a) Measured intensity profiles of CSAOVLs at different $z$ in a 2-$f$ lens system. (b) The corresponding measured phase distributions at the focal plane. (c) The calculated transverse energy flow at $z$=975 mm near the focal plane. Note that the values of $z$ and $n$ are labeled in the left side of each row and the top of each column, and the arrows in (c) are the directions of transverse energy flow. Other parameters in experiments are $a$ = 4 mm$^{-1}$, $b$ = 4. White scale bars in each row, 0.5 mm.

Figure 3 further elucidates the focusing characteristics of CSAOVLs with increasing $n$. It exhibits the intriguing and periodic intensity patterns of circularly-symmetric optical arrays with a period of 4 concerning the value of $n$. Such periodic distributions with $n$ is rooted in the similarity of the orbital-angular-momentum (OAM) spectra of these initial CSAOVLs with a period of 4. The detail mode purities of these CSAOVLs can be referred in Section D of the Supplemental Materials. The regular periodic variations of these focusing patterns are never noticed in the previous OVLs [39,41,42]. For example, for $n$ = 4$m$-3 ($m$ = 1, 2, 3, ⋯), as $m$ increases, the whole intensity profile remains essentially unchanged while the peripheral bright-spot structure becomes increasingly clear and evident. There are the same effects for other cases of $n = 4m$-2, $4m$-1 and $4m$. Notably, a distinctive bright spot appears in the central region when $n = 4m$, distinguishing them from arrays with other values of $n$. This bright spot in center has specific features shown in later. Furthermore, it is also noted that the value $n$ does not affect the rotational symmetry of the light field.

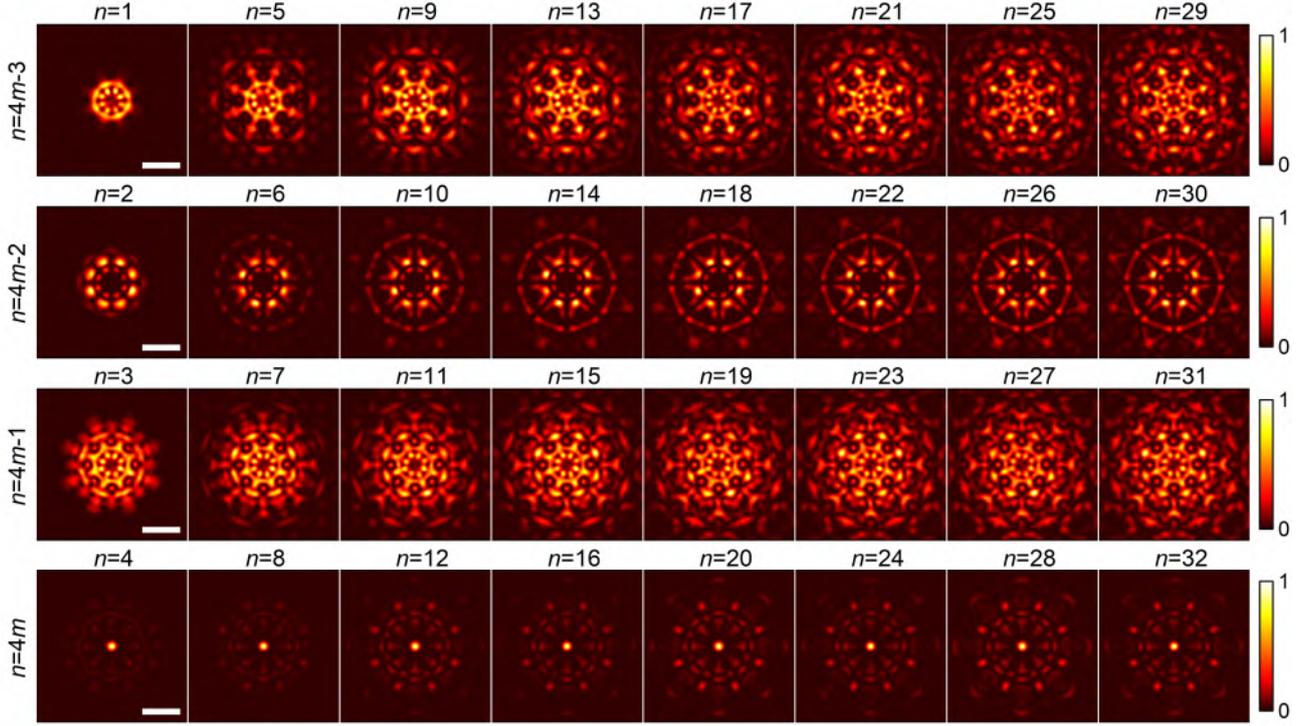

**Fig. 3.** Experimental intensity profiles of different-order CSAOVLs at the focal plane. Here, the top to bottom rows correspond to the different cases with $n = 4m$-3, $4m$-2, $4m$-1 and $4m$, respectively. Other parameters are the same as in Fig. 2. White scale bars, 0.5 mm.

Now, let us examine the impact of the radial and azimuthal modulation parameters ($a$ and $b$) on the resulting intensity distributions of these focusing CSAOVLs. Figure 4 shows the measured intensity distributions for four lowest-order CSAOVLs at the focal plane under varying $a$ and $b$. The experimental results agree with the theoretical predictions in <u>Section B</u> of the <u>Supplemental Materials</u>. In Fig. 4(a), in cases of small $a$, the light intensity patterns of the circular optical arrays concentrate in the center since the bright spots merge together. As $a$ increases, the circular optical arrays expand outward as a whole, resulting in a large- and rich-structure pattern. So, the role of $a$ serves as the radial control parameter for the size of the focused profiles. Also, the value of $a$ does not impact the rotational symmetry of the light field, since the parameter $a$ only affects slightly the weights of OAM spectra of CSAOVLs but does not change their components. For $n = 1$, 2 and 3, the central bright spot in the circular optical arrays gradually diminishes with increasing $a$. Particularly, for $n = 4$, the bright spot in the center always exists, thus a bright spot at the center is an intrinsic property for the CSAOVLs with $n = 4m$.

In Fig. 4(b), it further explores the effect of the azimuthal parameter $b$ on CSAOVLs. The intensity patterns of these CSAOVLs with $n = 1$, 2, 3 and 4 under different $b$ are experimentally measured. It shows that the number of bright spots increases along the azimuthal direction as $b$ increases. The value $b$ serves as a control for reshaping the focusing arrays along the azimuthal direction. When $b$ is odd, their patterns exhibit $b$-fold rotational symmetry. Conversely, if $b$ is even, the patterns display $2b$-fold rotational symmetry. Notably, a fractional value of $b$ can induce a more complex structure light field, contributing to a non-rotationally symmetrical structure of intensity patterns.

From the bottom row in Fig. 4(b), it is evident that a bright spot consistently exists in the center of circular optical arrays for $n = 4$, irrespective of variations in $b$. In contrast, for the cases of $n = 1$, 2 and 3 in Fig. 4(b), no bright spot appears in the center. Drawing insights from these observations in Fig. 4 and the periodic variation of intensity patterns in Fig. 3, the presence of a bright spot at the center can be affirmed as an intrinsic characteristic of circular optical arrays for $n = 4m$. The effects of both $a$ and $b$ on the OAM spectra and the influence of fractional $b$ on intensities can be seen in <u>Sections D and E</u> of the <u>Supplemental Materials.</u>

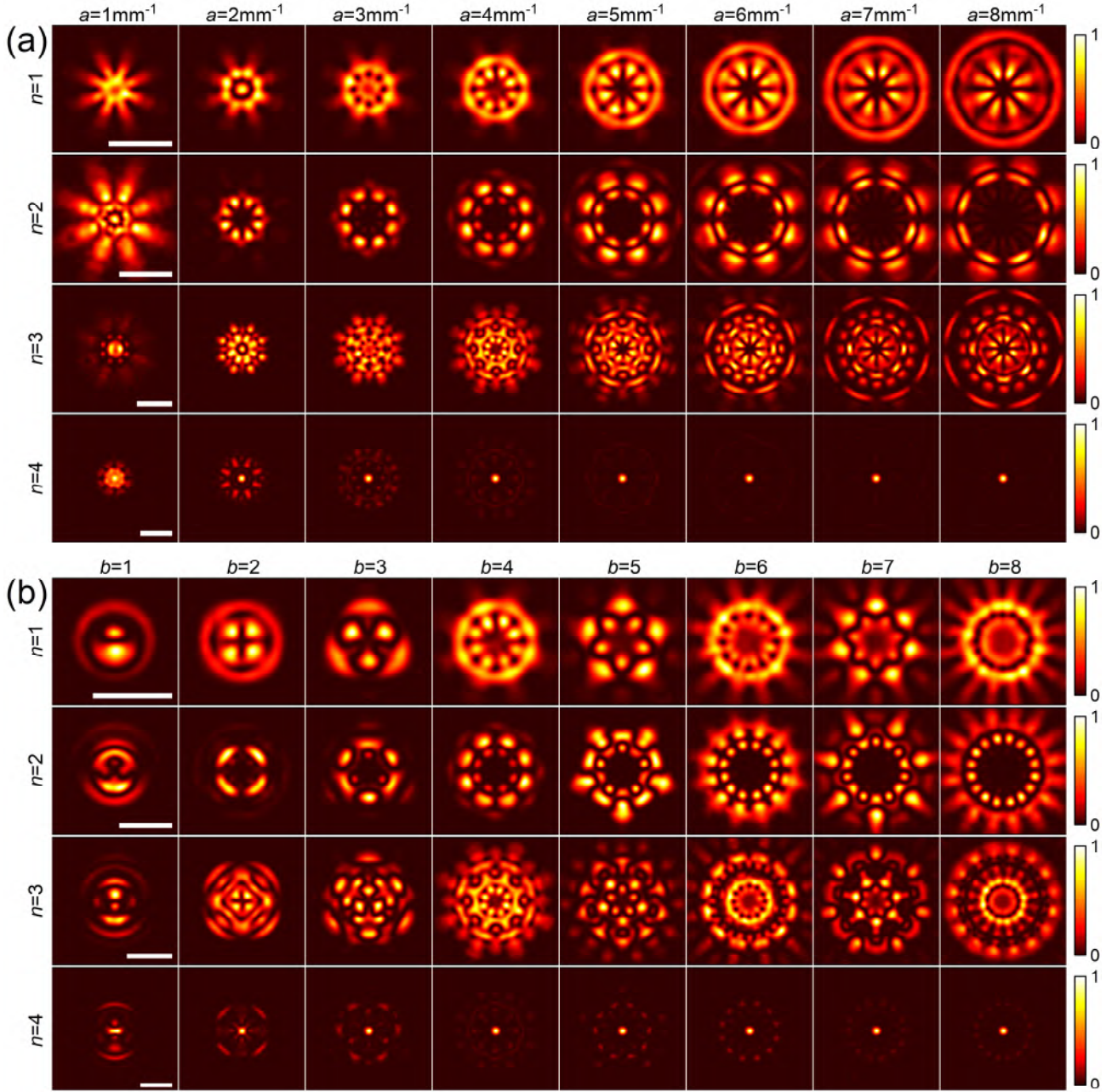

**Fig. 4.** Experimental intensity distributions of four lowest-order CSAOVLs at the focal plane under different radial and azimuthal modulations. In both (a) and (b), $n$=1, 2, 3, and 4 from the top to bottom rows, and both $a$ and $b$ are labeled on the top of each column. In (a), $b$ is fixed at $b = 4$, and in (b), $a$ is fixed at $a = 4$ mm$^{-1}$. Other parameters are the same as in Fig. 2. White scale bars, 0.5 mm.

It is noteworthy that the spot size plays a crucial role in a focused beam, since a smaller focusing spot size holds the potential to enhance spatial resolution of optical imaging. As concluded in the above, there always exists a center bright spot when $n$=4$m$, thus we experimentally conduct a qualitative analysis of the focusing spot sizes of these interesting CSAOVLs. Figure 5(a) shows the measured intensity distributions of the CSAOVLs with $n$=4$m$ at the focal plane. It shows that these focusing spot sizes is smaller than the GB under the same condition. For a clearer representation of the difference in focusing spot sizes between these CSAOVLs and the GB, the intensity profiles along both the $x$ and $y$ axes are presented in Figs. 5(b) and 5(c), respectively. These profiles confirm that the center bright spots of these CSAOVLs with $n = 4m$ are consistently smaller than the GB. Generally, a larger $n$ contributes to a smaller intensity profile, as distinctly demonstrated in the enlarging insets of Figs. 5(b) and 5(c). The corresponding theoretical results are provided in Section B of the Supplemental Materials.

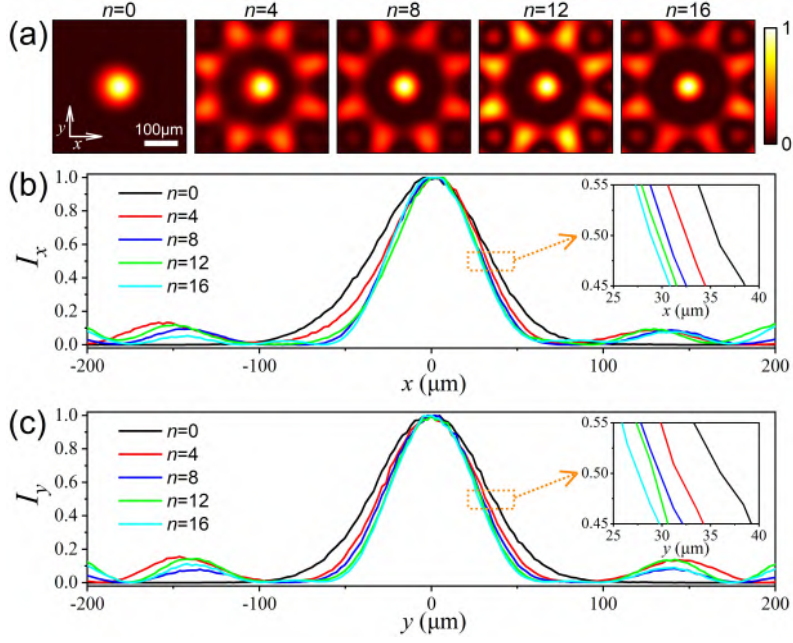

**Fig. 5.** Focusing properties of the CSAOVLs with $n = 4m$. (a) Experimental intensity distributions of the CSAOVLs with $n = 0, 4, 8,$ 12 and 16 at the focal plane. Note that the case of $n = 0$ corresponds to a GB (without modulation). (b)-(c) The intensity profiles of the focusing spots along the $x$ and $y$ axes, respectively. Other parameters are $a = 1.3$ mm$^{-1}$, $b = 4$.

**Table 1**

Comparison of the focal spot sizes between these CSAOVLs with $n = 4m$ and the GB (i.e. $n = 0$). $D_x$ and $D_y$ are the measured focal spot sizes in the $x$ and $y$ directions, respectively. The values in parentheses are the percentage reduction of the focal spot size for these CSAOVLs relative to the GB. Other parameters are the same as in Fig. 5.

|  | **$n=0$** | **$n=4$** | **$n=8$** | **$n=12$** | **$n=16$** |
|---|---|---|---|---|---|
| $D_x$ (μm) | 74.4 | 62.4 (16.1%) | 57.6 (22.6%) | 55.2 (25.8%) | 55.2 (25.8%) |
| $D_y$ (μm) | 72.0 | 62.4 (13.3%) | 57.6 (20.0%) | 55.2 (23.3%) | 52.8 (26.7%) |

Table 1 enumerates the experimentally-measured values of the focal spot sizes for these CSAOVLs with $n = 4m$ and the GB (i.e. $n = 0$). Here the focal spot size refers to the transverse full width at half-maximal intensity of a center bright spot. From Table 1, compared to the GB (i.e. $n = 0$), these CSAOVLs with $n = 4m$ exhibit smaller focal spot sizes in both the $x$ and $y$ directions. Notably, an increase in $n$ corresponds to a gradual decrease in focal-spot sizes in the $x$ and $y$ directions. For instance, when $n = 16$, the focal spot size of the CSAOVL is 25.8% smaller than that of the GB in the $x$ direction and 26.7% smaller in the $y$ direction. The results suggest the potential utility of such focusing circular optical arrays for the improvement of optical imaging. These measured results agree with the theoretical predictions, see in Section B of the Supplemental Materials.

## 4. Conclusions

In summary, we have experimentally generated arbitrary-order CSAOVL and revealed their propagation and focusing properties within a 2-$f$ focusing system or in the far-field region. The emergence of circular optical arrays is attributed to the merging and annihilation processes of vortex pairs with positive and negative TCs as approaching the focal plane. Notably, focused circular optical arrays exhibit the periodic patterns with respect to the value of $n$, which is in agreement with the periodicity in the OAM spectra of these CSAOVLs. The radial and azimuthal modulations offer flexible control over the focusing patterns of CSAOVLs. Particularly, their rotational symmetry can be modulated via the azimuthal parameter. When $n$ equals to multiples of 4, the size of the center bright spot occurring in the focusing patterns can be smaller than that of a GB. The distinctive properties of focused CSAOVLs may also hold promise for optical tests concerning the origin of OVLs in nematic liquid crystal light valves [51]. Finally, we anticipate that such circular optical arrays, composed of bright spots, will be useful in optical imaging and the exploration of light and micro-object interactions.

# Declaration of Competing Interest

The authors declare that they have no known competing financial interests or personal relationships that could have appeared to influence the work reported in this paper.

# Acknowledgements

This work was supported by the National Natural Science Foundation of China (grant No. 62375241).

# Appendix A. Supplementary material

Supplementary data associated with this article can be found, in the online version.

# References

[1] J.F. Nye, M.V. Berry, Dislocations in wave trains, Proc. R. Soc. A 336(1605) (1974) 165–190.

[2] P. Coullet, L. Gil, F. Rocca, Optical vortices, Opt. Commun. 73(5) (1989) 403-408.

[3] L. Allen, M.W. Beijersbergen, R.J.C. Spreeuw, J.P. Woerdman, Orbital angular momentum of light and the transformation of Laguerre-Gaussian laser modes, Phys. Rev. A 45(11) (1992) 8185–8189.

[4] D. Palacios, D. Rozas, G.A. Swartzlander, Observed scattering into a dark optical vortex core, Phys. Rev. Lett. 88(10) (2002) 103902.

[5] H. He, M.E.J. Friese, N.R. Heckenberg, H. Rubinszteindunlop, Direct observation of transfer of angular momentum to absorptive particles from a laser beam with a phase singularity, Phys. Rev. Lett. 75(5) (1995) 826-829.

[6] M. Padgett, R. Bowman, Tweezers with a twist, Nat. Photonics 5(6) (2011) 343-348.

[7] J. Wang, Advances in communications using optical vortices, Photonics Res. 4(5) (2016) B14-B28

[8] A. Vaziri, J.W. Pan, T. Jennewein, G. Weihs, A. Zeilinger, Concentration of higher dimensional entanglement: qutrits of photon orbital angular momentum, Phys. Rev. Lett. 91(22) (2003) 227902.

[9] X.L. Wang, Y.H. Luo, H.L. Huang, M.C. Chen, Z.E. Su, C. Liu, C. Chen, W. Li, Y.Q. Fang, X. Jiang, J. Zhang, L. Li, N.L. Liu, C.Y. Lu, J.W. Pan, 18-qubit entanglement with six photons' three degrees of freedom, Phys. Rev. Lett. 120(26) (2018) 260502.

[10] P.C. Maurer, J.R. Maze, P.L. Stanwix, L. Jiang, A.V. Gorshkov, A.A. Zibrov, B. Harke, J.S. Hodges, A.S. Zibrov, A. Yacoby, D. Twitchen, S.W. Hell, R.L. Walsworth, M.D. Lukin, Far-field optical imaging and manipulation of individual spins with nanoscale resolution, Nat. Phys. 6(11) (2010) 912-918.

[11] X.D. Qiu, F.S. Li, W.H. Zhang, Z.H. Zhu, L.X. Chen, Spiral phase contrast imaging in nonlinear optics: seeing phase objects using invisible illumination, Optica 5(2) (2018) 208-212.

[12] J.C. Ni, C.W. Wang, C.C. Zhang, Y.L. Hu, L. Yang, Z.X. Lao, B. Xu, J.W. Li, D. Wu, J.R. Chu, Three-dimensional chiral microstructures fabricated by structured optical vortices in isotropic material, Light. Sci. Appl. 6 (2017) e17011.

[13] L.H. Zhu, M.M. Tang, H.H. Li, Y.P. Tai, X.Z. Li, Optical vortex lattice: an exploitation of orbital angular momentum, Nanophotonics 10(9) (2021) 2487-2496.

[14] X. Li, Y. Zhou, Y.N. Cai, Y.N. Zhang, S.H. Yan, M.M. Li, R.Z. Li, B.L. Yao, Generation of hybrid optical trap array by holographic optical tweezers, Front. Physics 9 (2021) 591747.

[15] Z.S. Wan, Y.J. Shen, Z.Y. Wang, Z.J. Shi, Q. Liu, X. Fu, Divergence-degenerate spatial multiplexing towards future ultrahigh capacity, low error-rate optical communications, Light. Sci. Appl. 11(1) (2022) 144.

[16] K. Ladavac, D.G. Grier, Microoptomechanical pumps assembled and driven by holographic optical vortex arrays, Opt. Express 12(6) (2004) 1144-1149.

[17] A. Mair, A. Vaziri, G. Weihs, A. Zeilinger, Entanglement of the orbital angular momentum states of photons, Nature 412(6844) (2001) 313-316.

[18] S. Franke-Arnold, J. Leach, M.J. Padgett, V.E. Lembessis, D. Ellinas, A.J. Wright, J.M. Girkin, P. Öhberg, A.S. Arnold, Optical ferris wheel for ultracold atoms, Opt. Express 15(14) (2007) 8619-8625.

[19] A.S. Arnold, Extending dark optical trapping geometries, Opt. Lett. 37(13) (2012) 2505-2507.

[20] T. Lei, M. Zhang, Y.R. Li, P. Jia, G.N. Liu, X.G. Xu, Z.H. Li, C.J. Min, J. Lin, C.Y. Yu, H.B. Niu, X.C. Yuan, Massive individual orbital angular momentum channels for multiplexing enabled by Dammann gratings, Light. Sci. Appl. 4 (2015) e257.

[21] J.P. Liu, C.J. Min, T. Lei, L.P. Du, Y.S. Yuan, S.B. Wei, Y.P. Wang, X.C. Yuan, Generation and detection of broadband multi-channel orbital angular momentum by micrometer-scale meta-reflectarray, Opt. Express 24(1) (2016) 212-218.

[22] G.X. Wang, X.Y. Kang, X.J. Sun, Z.Y. Li, Y. Li, K.Y. Chen, N. Zhang, X.M. Gao, S.L. Zhuang, Generation of perfect optical vortex arrays by an optical pen, Opt. Express 30(18) (2022) 31959-31970.

[23] A.M. Khazaei, D. Hebri, S. Rasouli, Theory and generation of heterogeneous 2D arrays of optical vortices by using 2D fork-shaped gratings: topological charge and power sharing management, Opt. Express 31(10) (2023) 16361-16379.

[24] C. Xu, X. Chen, Y.Y. Cai, Y.Q. Wang, High-quality tunable optical vortex arrays with multiple states of orbital angular momentum, Opt. Laser Technol. 169 (2024) 110029.

[25] K.B. Yang, H. Luo, Y.D. Zhang, P. Li, F. Wen, Y.Z. Gu, Z.K. Wu, Modulating and identifying an arbitrary curvilinear phased optical vortex array of high-order orbital angular momentum, Opt. Laser Technol. 168 (2024) 109984.

[26] C.J. Min, J.P. Liu, T. Lei, G.Y. Si, Z.W. Xie, J. Lin, L.P. Du, X.C. Yuan, Plasmonic nano-slits assisted polarization selective detour phase meta-hologram, Laser Photon. Rev. 10(6) (2016) 978-985.

[27] H. Gao, Y. Li, L.W. Chen, J.J. Jin, M.B. Pu, X. Li, P. Gao, C.T. Wang, X.G. Luo, M.H. Hong, Quasi-Talbot effect of orbital angular momentum beams for generation of optical vortex arrays by multiplexing metasurface design, Nanoscale 10(2) (2018) 666-671.

[28] E.L. Wang, L.N. Shi, J.B. Niu, Y.L. Hua, H.L. Li, X.L. Zhu, C.Q. Xie, T.C. Ye, Multichannel spatially nonhomogeneous focused vector vortex beams for

quantum experiments, Adv. Opt. Mater. 7(8) (2019) 1801415.

[29] E. Brasselet, Tunable optical vortex arrays from a single nematic topological defect, Phys. Rev. Lett. 108(8) (2012) 087801.

[30] B. Son, S. Kim, Y.H. Kim, K. Käläntär, H.M. Kim, H.S. Jeong, S.Q. Choi, J. Shin, H.T. Jung, Y.H. Lee, Optical vortex arrays from smectic liquid crystals, Opt. Express 22(4) (2014) 4699-4704.

[31] P. Salamon, N. Éber, Y. Sasaki, H. Orihara, A. Buka, F. Araoka, Tunable optical vortices generated by self-assembled defect structures in nematics, Phys. Rev. Appl. 10(4) (2018) 044008.

[32] P. Chen, L.L. Ma, W. Duan, J. Chen, S.J. Ge, Z.H. Zhu, M.J. Tang, R. Xu, W. Gao, T. Li, W. Hu, Y.Q. Lu, Digitalizing self-assembled chiral superstructures for optical vortex processing, Adv. Mater. 30(10) (2018) 1705865.

[33] Z.H. Li, C.F. Cheng, Generation of second-order vortex arrays with six-pinhole interferometers under plane wave illumination, Appl. Optics 53(8) (2014) 1629-1635.

[34] S. Vyas, P. Senthilkumaran, Vortex array generation by interference of spherical waves, Appl. Optics 46(32) (2007) 7862-7867.

[35] S.C. Chu, C.S. Yang, K. Otsuka, Vortex array laser beam generation from a Dove prism-embedded unbalanced Mach-Zehnder interferometer, Opt. Express 16(24) (2008) 19934-19949.

[36] A. Dudley, A. Forbes, From stationary annular rings to rotating Bessel beams, J. Opt. Soc. Am. A 29(4) (2012) 567-573.

[37] H.X. Ma, X.Z. Li, Y.P. Tai, H.H. Li, J.G. Wang, M.M. Tang, J. Tang, Y.S. Wang, Z.G. Nie, Generation of circular optical vortex array, Ann. Phys. (Berlin) 529(12) (2017) 1700285.

[38] X.D. Qiu, F.S. Li, H.G. Liu, X.G. Chen, L.X. Chen, Optical vortex copier and regenerator in the Fourier domain, Photonics Res. 6(6) (2018) 641-646.

[39] L. Stoyanov, G. Maleshkov, M. Zhekova, I. Stefanov, D.N. Neshev, G.G. Paulus, A. Dreischuh, Far-field pattern formation by manipulating the topological charges of square-shaped optical vortex lattices, J. Opt. Soc. Am. B 35(2) (2018) 402-409.

[40] L. Zhu, A.D. Wang, M.L. Deng, B. Lu, X.J. Guo, Experimental demonstration of multiple dimensional coding decoding for image transfer with controllable vortex arrays, Sci Rep 11(1) (2021) 12012.

[41] D.D. Liu, B.J. Gao, F.J. Wang, J.S. Wen, L.G. Wang, Experimental realization of tunable finite square optical arrays, Opt. Laser Technol. 153 (2022) 108220.

[42] L. Stoyanov, G. Maleshkov, M. Zhekova, I. Stefanov, G.G. Paulus, A. Dreischuh, Far-field beam reshaping by manipulating the topological charges of hexagonal optical vortex lattices, J. Opt. 20(9) (2018) 095601.

[43] H.H. Fan, H. Zhang, C.Y. Cai, M.M. Tang, H.H. Li, J. Tang, X.Z. Li, Flower-shaped optical vortex array, Ann. Phys. (Berlin) 533(4) (2021) 2000575.

[44] H. Wang, S.Y. Fu, C.Q. Gao, Tailoring a complex perfect optical vortex array with multiple selective degrees of freedom, Opt. Express 29(7) (2021) 10811-10824.

[45] L. Li, C.L. Chang, X.Z. Yuan, C.J. Yuan, S.T. Feng, S.P. Nie, J.P. Ding, Generation of optical vortex array along arbitrary curvilinear arrangement, Opt. Express 26(8) (2018) 9798-9812.

[46] S.A. Collins, Lens-system diffraction integral written in terms of matrix optics, J. Opt. Soc. Am. 60 (9) (1970) 1168 –1177.

[47] S. Wang, D. Zhao, Matrix optics, CHEP-Springer (2000).

[48] S. Wen, L.G. Wang, X.H. Yang, J.X. Zhang, S.Y. Zhu, Vortex strength and beam propagation factor of fractional vortex beams, Opt. Express 27 (4) (2019) 5893–5904.

[49] J.A. Davis, D.M. Cottrell, J. Campos, M.J. Yzuel, I. Moreno, Encoding amplitude information onto phase-only filters, Appl. Opt. 38 (23) (1999) 5004 –5013.

[50] M. Takeda, H. Ina, S. Kobayashi, Fourier-transform method of fringe-pattern analysis for computer-based topography and interferometry, J. Opt. Soc. Am. 72(1) (1982) 156-160.

[51] E. Calisto, M.G. Clerc, M. Kowalczyk, P. Smyrnelis, On the origin of the optical vortex lattices in a nematic liquid crystal light valve, Opt. Lett. 44 (12) (2019) 2947–2950.

# Supplementary Materials
*of*
# Circularly-Symmetric Alternating Optical Vortex Lattices and their Focusing Characteristics

## Dadong Liu and Li-Gang Wang*

*School of Physics, Zhejiang University, Hangzhou, 310058, China*
*E-mail address: lgwang@zju.edu.cn*

## A.  Formulae of light fields in paraxial linear optical systems

Based on the paraxial approximation, the output light field through a linear optical system can be theoretically predicted by the Collins formula [1,2]

$$E_{\mathrm{o}}^{(n)}(\rho,\theta,z)=\frac{\exp(ikz)}{i\lambda B}\iint E_{\mathrm{i}}^{(n)}(r,\varphi)\exp\{\frac{ik}{2B}[Ar^2-2r\rho\cos(\theta-\varphi)+D\rho^2]\}r\mathrm{d}r\mathrm{d}\varphi, \qquad (S1)$$

where $E_{\mathrm{i}}^{(n)}(r,\varphi)$ and $E_{\mathrm{o}}^{(n)}(\rho,\theta,z)$ are, respectively, the input and output fields at the initial and observation planes, $A$, $B$, and $D$ are the elements of a ray transfer matrix $\begin{pmatrix} A & B \\ C & D \end{pmatrix}$ describing a linear optical system, $\lambda$ is the wavelength of light, $k=2\pi/\lambda$ is the wave number, and $z$ is the distance along the propagation axis from the initial plane to the output (or observation) plane.

The intensity distributions of these output fields are calculated via $I_{\mathrm{o}}^{(n)}(\rho,\theta,z)=\left|E_{\mathrm{o}}^{(n)}(\rho,\theta,z)\right|^2$. For the fields propagating inside a 2-*f* lens system, its ray transfer matrix between the initial (or generation) and observation planes is given by [2]

$$\begin{pmatrix} A & B \\ C & D \end{pmatrix}=\begin{pmatrix} 1-(z-f)/f & f \\ -1/f & 0 \end{pmatrix}, \qquad (S2)$$

where *f* is the focal length of the 2-*f* lens system, and *z* is the propagation distance from the generation plane as shown in Fig.1(c). Specifically, when $z=2f$ (*i.e.*, at the focal plane of the 2-*f* lens system), its ray transfer matrix simplifies to [3]

$$\begin{pmatrix} A & B \\ C & D \end{pmatrix}=\begin{pmatrix} 0 & f \\ -1/f & 0 \end{pmatrix}. \qquad (S3)$$

In this situation, Eq. (S1) is equivalent to the Fraunhofer diffraction and the focusing fields correspond to the cases of the far-field regions of free space. Based on Eq. (1) in manuscript and Eq. (S1) and the ray matrix (S2), the intensity evolutions of such circularly-symmetric alternating optical vortex lattices (CSAOVLs) at any propagation distance *z* of the 2-*f* lens system can be theoretically predicted. Notably, it is well known that the propagation of light in the 2-*f* lens system from the lens to the rear focal plane is analogous to the propagation of light in free space from the near-field to far-field regions. This consequently offers greater convenience for experimental investigation on the intensity evolution of such CSAOVLs in free space and their focusing features.

## B.  Theoretical intensity distributions of the CSAOVLs in a 2-*f* lens system under different parameters *a*, *b*, *n* and *z*

In this section, we present all theoretical predictions that correspond to the experimental figures used in the discussion. Fig. S1 shows more experimental intensity evolutions of four lowest-order CSAOVLs at different propagation distances *z* in a 2-*f* lens system. Fig. S2 shows the corresponding theoretical intensity evolutions of these four lowest-order CSAOVLs in Fig. S1. These results show the agreement of experimental and theoretical intensity evolutions at different propagation distances under different order numbers of *n*.

Figure S3 demonstrate the corresponding theoretical evolutions of phase and transverse energy flow for four lowest-order CSAOVLs approaching the focal plane. From Fig. S3(a), we can see the evolution of phase singularities: vortex nucleation, evolution, and annihilation, and during these complex dynamics there are also the formation of fast-changing π-phase-like edge dislocations. These phase singularities and phase edge dislocations correspond to the dark-field structures and form the patterns of bright spots at the focal plane. In Fig. S3(b), it shows how the transverse energy flows of these CSAOVLs evolve as they approach the focal plane. It is seen that, for the case of $n=4$ the transverse energy flows near the center region are always towards the center before the focal plane, while for other three cases of $n=1$, 2, and 3, the transverse energy flows in the center region will be directed outward from the inner region when they close to the focal plane. Thus, there is a dark region in the center region for other three cases of $n=1$, 2, and 3 at the focal plane while for $n=4$ there is a bright spot in the center at the focal plane.

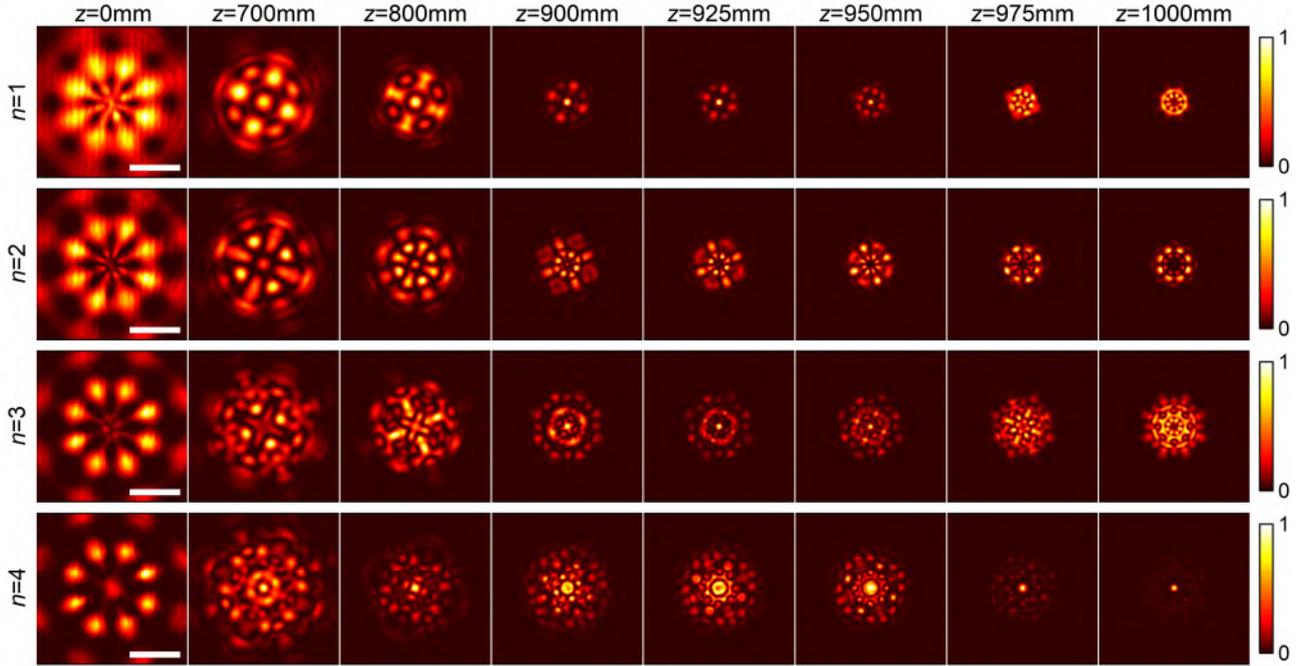

**Fig. S1.** Complete experimental intensity evolutions of four lowest-order CSAOVLs at different propagation distances $z$ in a 2-$f$ lens system. Here, the values of $n = 1$, 2, 3, and 4 are noted in the left side of each row, and the distances of $z=0$, 700mm, 800mm, 900mm, 925mm, 950mm, 975mm, and 1000mm are labeled on the top of each column. White scale bars, 1 mm.

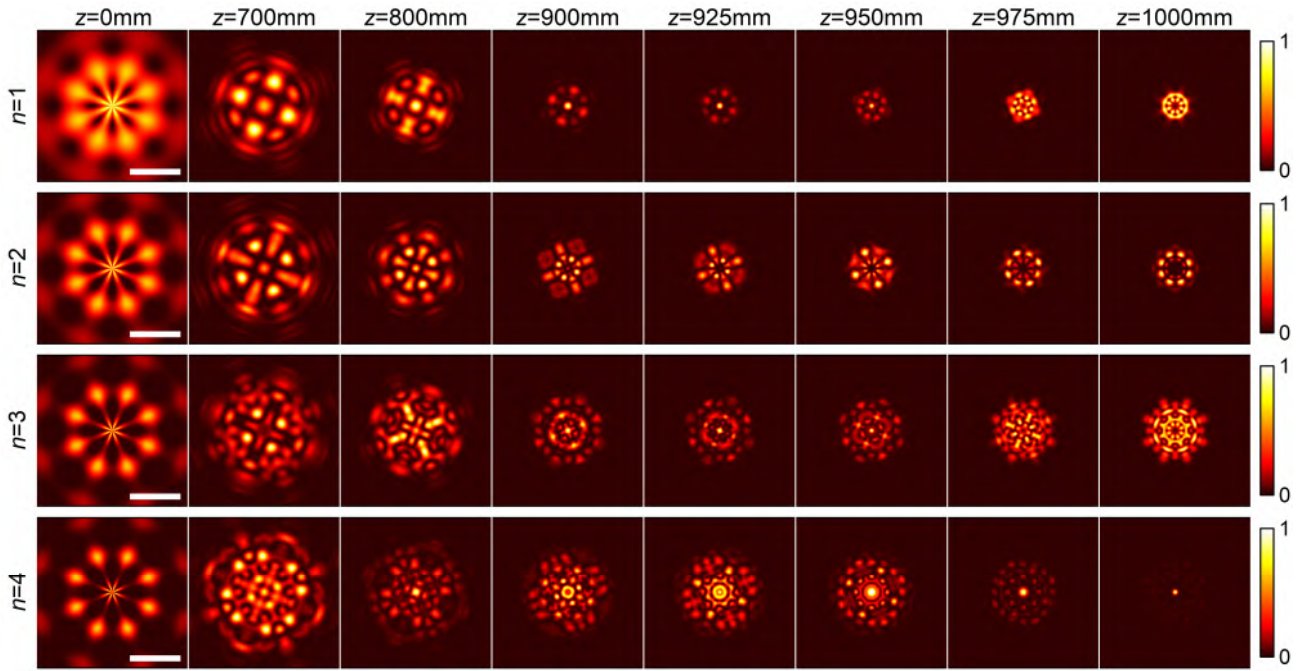

**Fig. S2.** The corresponding theoretical intensity evolutions of four lowest-order CSAOVLs of Fig. S1. White scale bars, 1 mm.

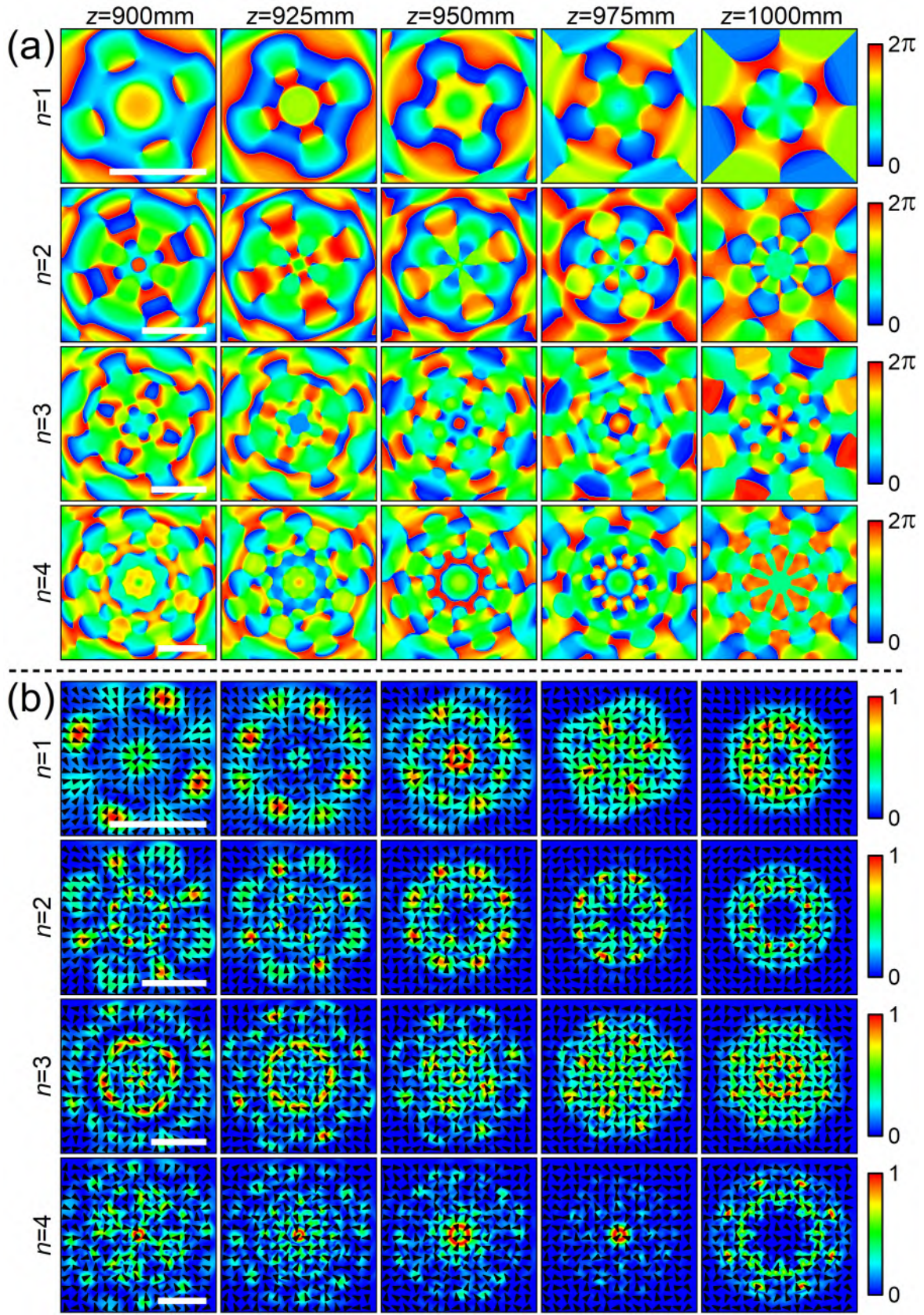

**Fig. S3.** The corresponding theoretical evolutions of phase and transverse energy flow for four lowest-order CSAOVLs with $n$ = 1, 2, 3, and 4, closing to the focal plane. Here, $z$=900 mm, 925 mm, 950 mm, 975 mm, and 1000 mm. White scale bars, 0.5 mm.

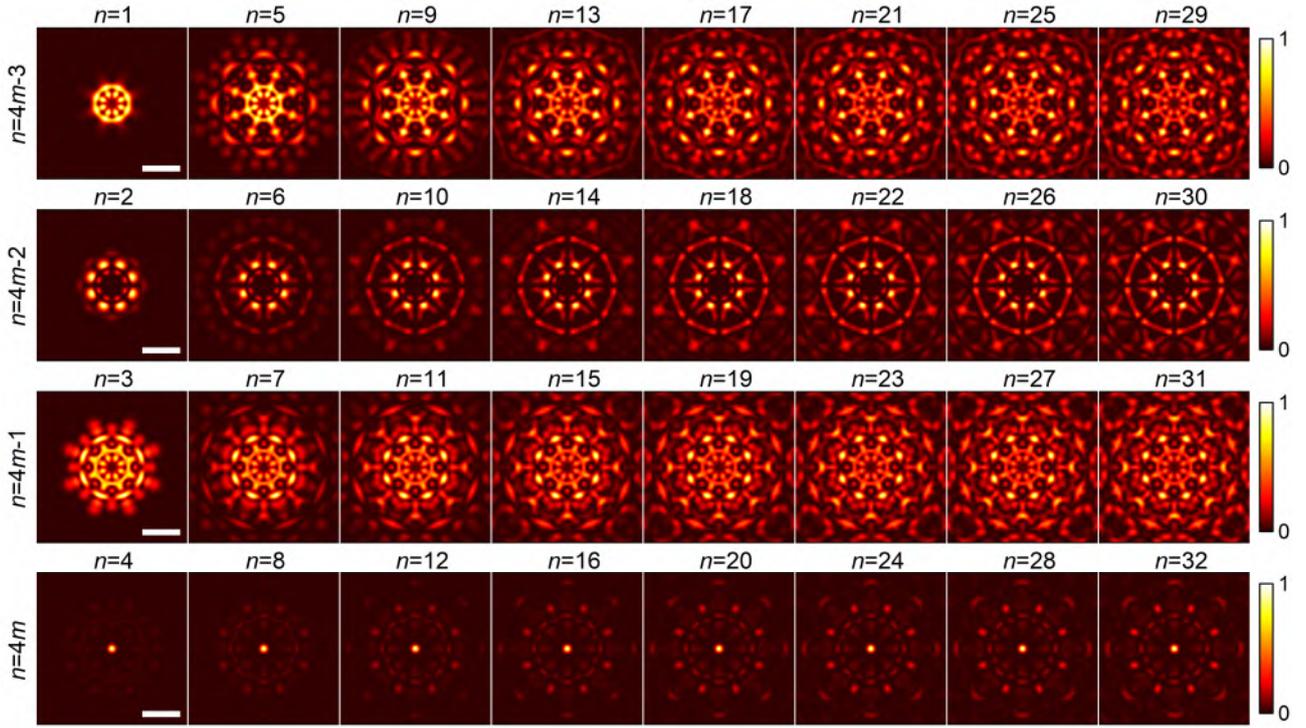

**Fig. S4**. The corresponding theoretical intensity profiles of different-order CSAOVLs at the focal plane. Here, from the top to bottom columns they correspond to the different cases with $n = 4m$-3, $4m$-2, $4m$-1 and $4m$, respectively. The parameters are $a = 4$ mm$^{-1}$, $b = 4$, $w_0 = 1.63$ mm, and $f = 500$ mm. White scale bars, 0.5 mm.

Figure S4 shows the theoretical intensity profiles of different-order CSAOVLs at the focal plane, corresponding to the situations of the experimental results in Fig. 3. Clearly, these theoretical intensity profiles do well predict the experimental observations.

Figure S5 shows the theoretical intensity profiles of four lowest-order CSAOVLs at the focal plane under different radial and azimuthal modulations. These results do well explain our experimental measurements.

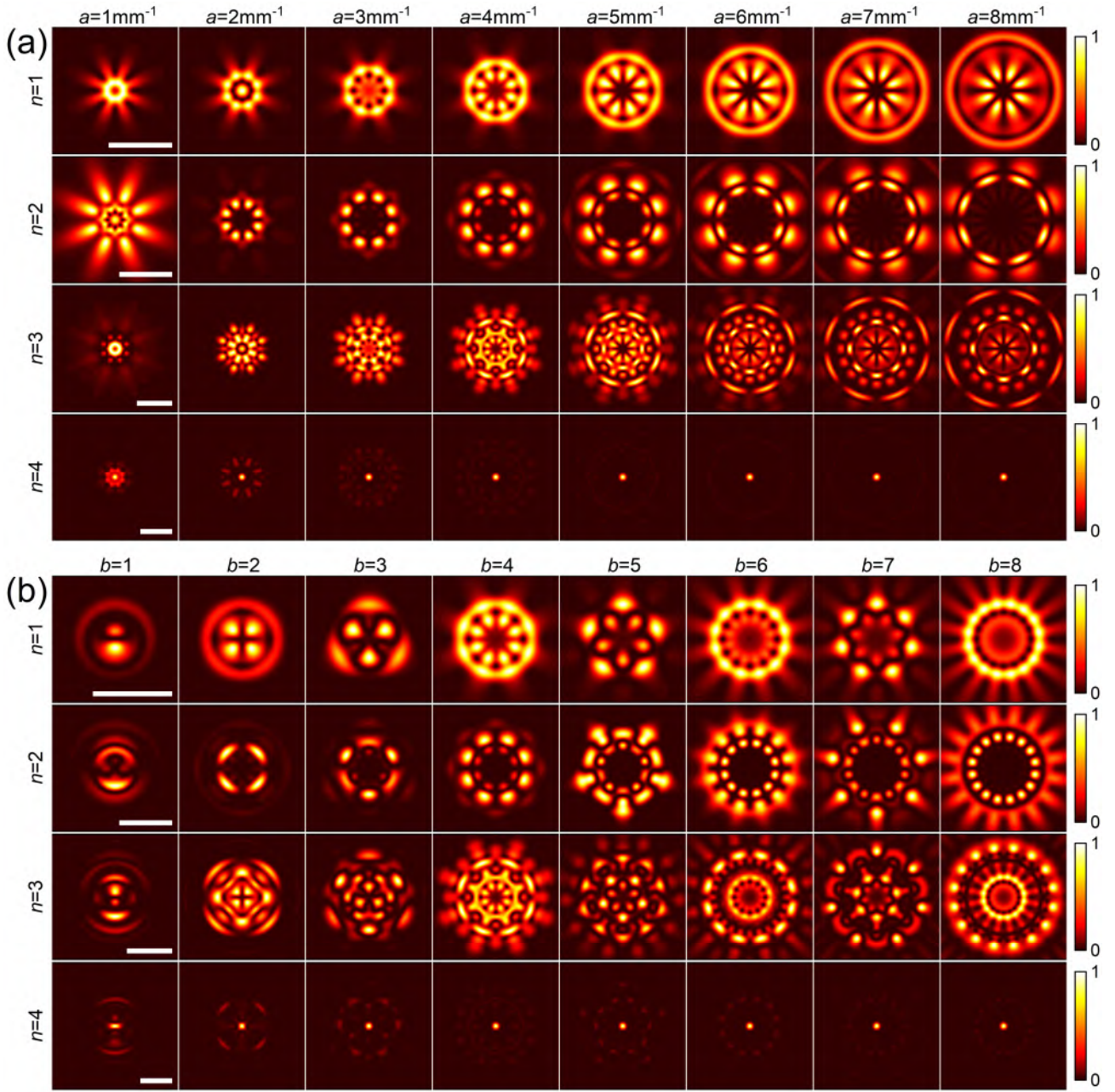

**Fig. S5.** Theoretical intensity distributions of four lowest-order CSAOVLs at the focal plane under different radial and azimuthal modulations. In both (a) and (b), $n$=1, 2, 3, and 4 from the top to bottom rows, and both $a$ and $b$ change and are labeled on the top of each column. In (a), $b$ is fixed at $b = 4$, and in (b), $a$ is fixed at $a = 4$ mm$^{-1}$. Other parameters are $w_0 = 1.63$ mm and $f = 500$ mm. White scale bars, 0.5 mm.

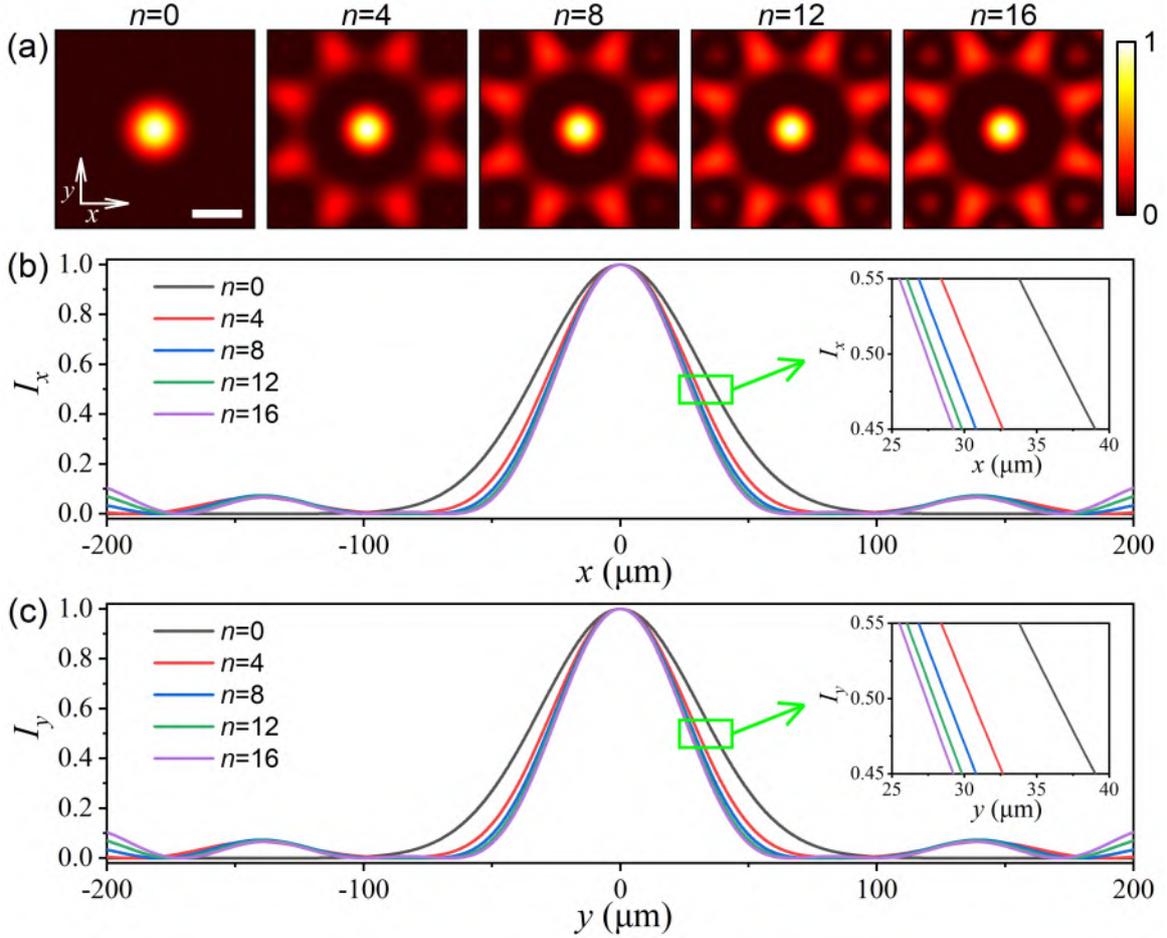

**Fig. S6.** Focusing properties of the CSAOVLs with $n = 4m$. (a) Theoretical intensity distributions of the CSAOVLs with $n = 0$, 4, 8, 12 and 16 at the focal plane. Note that the case of $n = 0$ corresponds to a Gaussian beam (without modulation). (b)-(c) The intensity profiles of the focusing spots along the $x$ and $y$ axes, respectively. Other parameters are $a = 1.3$ mm$^{-1}$, $b = 4$, $w_0 = 1.63$ mm, and $f = 500$ mm. The white scale bar denotes 100 μm.

**Table S1**

Comparison of the focal spot sizes between these CSAOVLs with $n = 4m$ and the Gaussian beam (i.e. $n = 0$). $D_x$ and $D_y$ are the calculated focal spot sizes in the $x$ and $y$ directions, respectively. The values in parentheses are the percentage reduction of the focal spot size for these CSAOVLs relative to the Gaussian beam. Other parameters are the same as in Fig. S6.

|  | $n=0$ | $n=4$ | $n=8$ | $n=12$ | $n=16$ |
|---|---|---|---|---|---|
| $D_x$ (μm) | 72.8 | 60.8 (16.5%) | 57.6 (20.9%) | 56.0 (23.1%) | 54.4 (25.3%) |
| $D_y$ (μm) | 72.8 | 60.8 (16.5%) | 57.6 (20.9%) | 56.0 (23.1%) | 54.4 (25.3%) |

Figure S6 plots the theoretical predictions of Fig. 5, and Table S1 presents the theoretical values of Table 1. All these results show that our experimental results agree well with these theoretical predications.

## C.   Theoretical evolution of vortices for the CSAOVLs with different $n$ in a 2-$f$ lens system

The progression of vortices within CSAOVLs is explored in Fig. S7, elucidating their evolution from the lens to the focal plane. Initially, as $z$ extends from the initial plane toward the lens, numerous pairs of positive and negative vortices are observed. This observation aligns with the composition of the initial fields, characterized by a succession of alternating OVs. Subsequently, beyond the lens, with an increase in $z$, positive and negative OVs systematically converge and annihilate. In Fig. S7(a1), corresponding to $n = 1$, pairs of positive and negative vortices are discernible near the lens plane. As $z$ increases, nearing the focal plane, trajectories of these alternated OVs intersect, indicative of the annihilation processes. For larger $n$ values (i.e. $n = 2$, 3 and 4), as illustrated in Figs. S7(b1)-S7(d1), vortex trajectories become denser, signifying more intricate annihilation dynamics. Despite the complexity in the evolution of optical vortices, the trend is towards mutual annihilation during propagation. Enlarging views of Figs. S7(a2)-S7(d2) offer clearer insights into the generation and annihilation processes of OVs. In the focal plane, a series of persistent vortices emerge, contributing to the formation of dark regions amid bright spots within circular optical arrays. Consequently, the annihilation of vortices plays a pivotal role in shaping circular optical arrays, while the presence of surviving vortices influences the intensity distributions of these arrays. Additionally, trajectories of these OVs in CSAOVLs possess four-fold rotational symmetry, as evident in their projections in the $x$-$y$ plane (depicted by green lines) in Figs. S7(a1)-S7(d1).

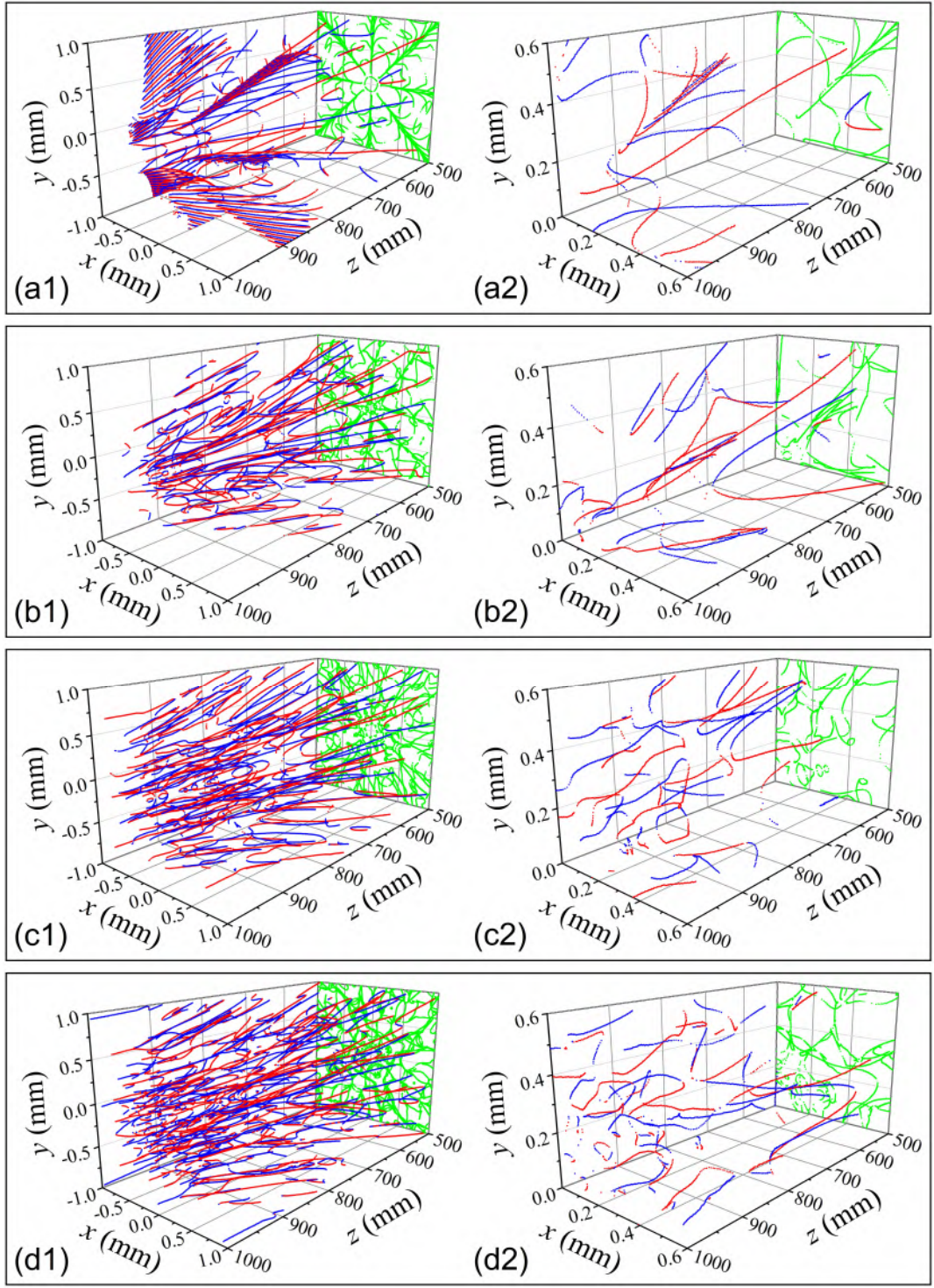

**Fig. S7.** Evolution of vortices of the CSAOVLs with (a1) $n = 1$, (b1) $n = 2$, (c1) $n = 3$ and (d1) $n = 4$, approaching the focal plane of the 2-$f$ lens system. (a2)-(d2) are the enlarging parts in (a1)-(d1). The blue and red lines (dots) denote, respectively, the trajectories of positive and negative vortices, and their projections are presented by the green lines (dots) in the x-y plane. Other parameters are $a = 4$ mm$^{-1}$, $b = 4$, $w_0 = 1.63$ mm, and $f = 500$ mm.

## D. Orbital angular momentum (OAM) spectra of the CSAOVLs with different *a*, *b* and *n* at the initial plane

The helical harmonic exp($il\varphi$) serves as the eigenfunction of orbital angular momentum (OAM), where $l$ denotes the topological charge (TC). A light field $E(r,\varphi,z)$ can be decomposed into helical harmonics exp($il\varphi$) as [4,5]

$$E(r,\varphi,z) = \frac{1}{\sqrt{2\pi}} \sum_{l=-\infty}^{+\infty} a_l(r,z) \exp(il\varphi), \tag{S4}$$

where the coefficients $a_l(r,z)$ can be expressed as

$$a_l(r,z) = \frac{1}{\sqrt{2\pi}} \int_0^{2\pi} E(r,\varphi,z) \exp(-il\varphi) \mathrm{d}\varphi. \tag{S5}$$

The intensity of the $l$-th order helical harmonic, independent of the parameter $z$, can be defined as

$$C_l = \int_0^{+\infty} |a_l(r,z)|^2 r \mathrm{d}r. \tag{S6}$$

The intensity weight of the helical harmonic is given by

$$R_l = \frac{C_l}{\sum_{q=-\infty}^{+\infty} C_q}, \tag{S7}$$

which can be regard as the OAM spectra or mode purities of the light field $E(r,\varphi,z)$ [6,7].

Figure S8 shows the OAM spectra of the CSAOVLs under different $n$ at the initial plane. It is observed that the intensity weights of the OAM spectra for the CSAOVLs exhibit symmetry about TC $l = 0$, consistent with the alternated optical vortex lattices' nature, possessing positive and negative TCs $\pm n$ in equal proportions. Notably, the OAM spectra of CSAOVLs demonstrate a periodicity of 4 concerning the parameter $n$. Specifically, the intensity weights of the OAM spectra can be categorized into four groups based on the parameter $n$. For cases where $n = 4m-3$ with $m = 1, 2, 3, \cdots$, being positive integers, the OAM spectra patterns of CSAOVLs expand based on that for $m= 1$ (i.e., $n = 1$). Similarly, for cases of $n = 4m-2, 4m-1$, and $4m$, the OAM spectra patterns also expand relative to $m = 1$ (i.e., $n = 2, 3$, and 4). This periodic variation in the OAM spectra of CSAOVLs with respect to the parameter $n$ mirrors the periodic changes observed in the intensity patterns of circular optical arrays.

Figure S9 presents the influences of the parameter $n$ on the number of OAM modes for CSAOVLs, displayed in logarithmic scales for clarity. The analysis reveals that the number of OAM modes increases proportionally with the parameter $n$. Precisely, the total number $N$ of OAM modes is given by the function $N = 2n + 1$. For instance, when $n$ takes values of 1, 2, 3, and 4, the corresponding numbers $N$ of OAM modes are 3, 5, 7, and 9, respectively. Therefore, the parameter $n$ serves as a means to regulate the quantity of OAM modes inherent in the CSAOVLs.

Figure S10 demonstrates the impact of the parameter $a$ on the intensity weights of OAM modes for CSAOVLs with $n = 1$. The analysis reveals a close correlation between the intensity weights of OAM modes and the parameter $a$. Specifically, for $n = 1$, as the value of $a$ increases, the intensity weight of the zero-order OAM mode gradually rises, while the intensity weights of positive and negative fourth-order OAM modes progressively decrease, as depicted in Fig. S10. Although the evolution of intensity weights for OAM modes becomes more intricate for larger $n$, the parameter $a$ exclusively influences the intensity weights of OAM modes. Consequently, the parameter $a$ can be used as a control parameter for the redistribution of intensity weights among OAM modes.

Figures S11 and S12 show the influences of the parameter $b$ on the spacing of OAM modes for CSAOVLs with $n = 1$. As depicted in Fig. S11, when $b$ is a positive integer, the spacing between adjacent OAM modes remains constant and is equivalent to the value of $b$. It is essential to note that these findings are applicable across all values of $n$ and are not confined to $n = 1$. For instance, when $b = 4$, the spacing between adjacent OAM modes for CSAOVLs with $n$ varying from 1 to 12 remains constant at 4 (i.e., the value of $b$), as illustrated in Fig. S9. The integer values of $b$ solely impact the spacing between adjacent OAM modes without affecting the intensity weights of OAM modes, as clearly illustrated in Fig. S11. Furthermore, the influence of fractional $b$ on the OAM spectra is multifaceted, as it can alter the intensity weights of OAM modes and increase the number of OAM modes, as observed in Fig. S12.

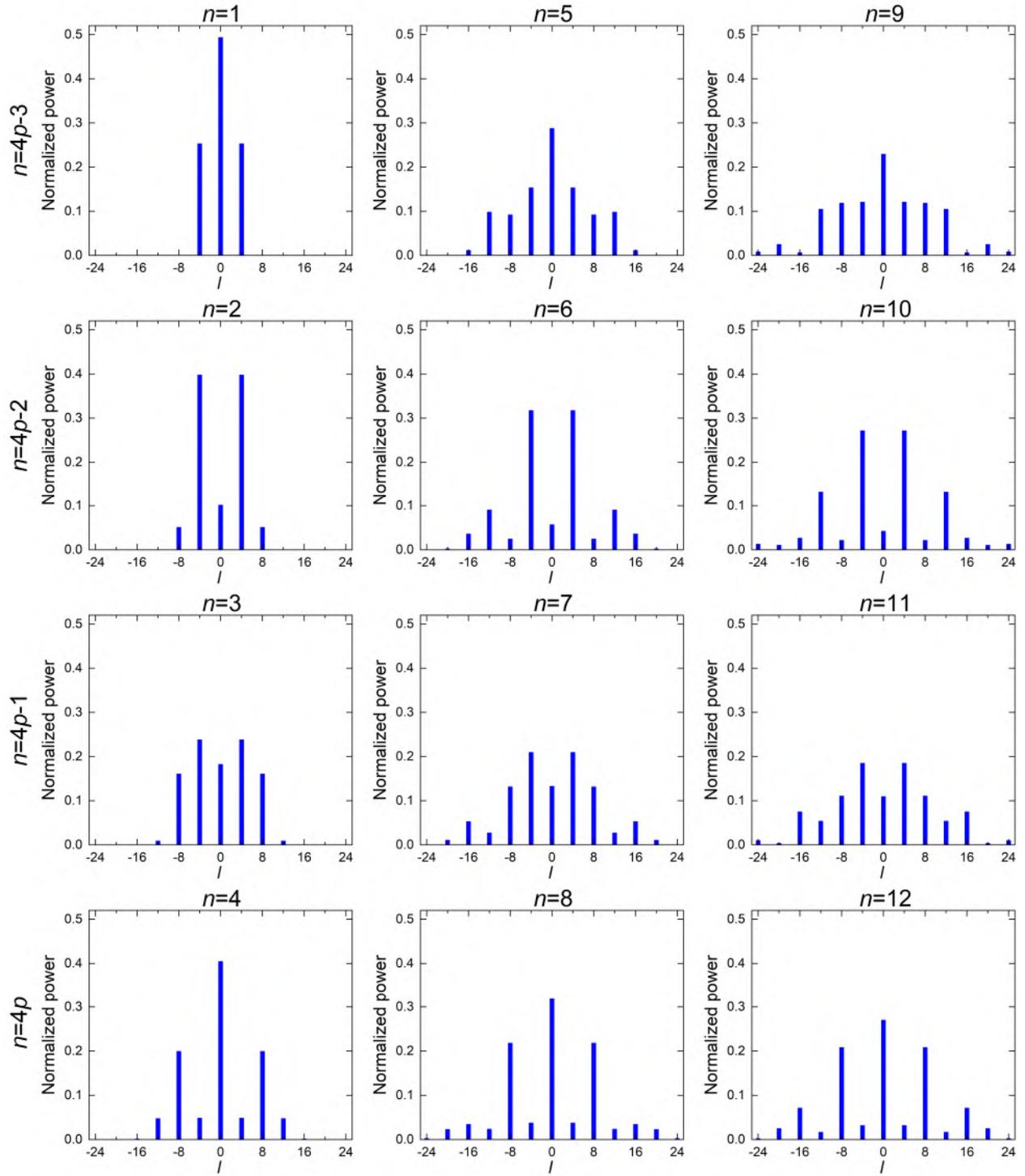

**Fig. S8.** OAM spectra of the CSAOVLs with different $n$ at the initial plane. Other parameters are $a$ = 4 mm$^{-1}$, $b$ = 4, and $w_0$ = 1.63 mm.

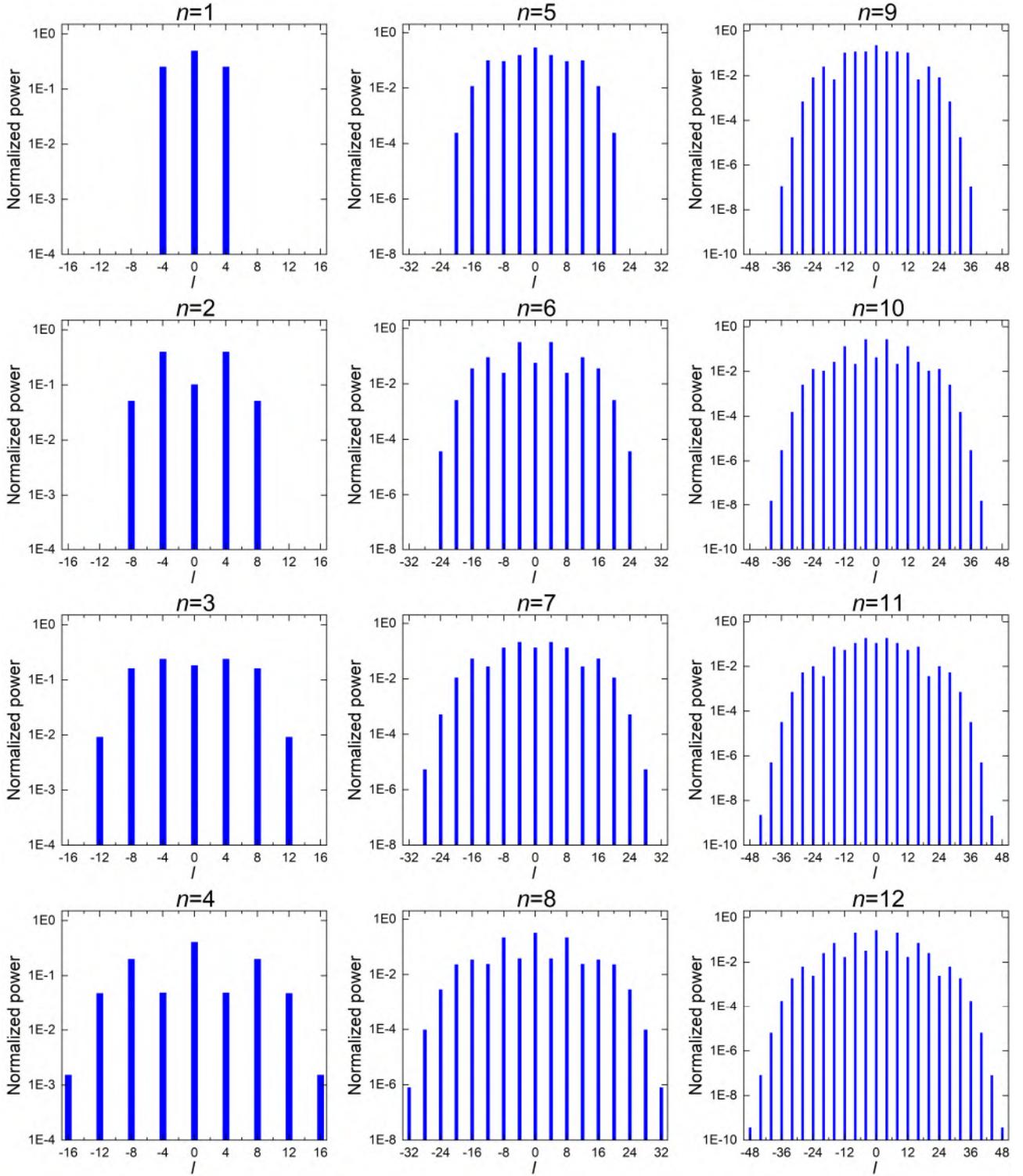

**Fig. S9.** OAM spectra of the CSAOVLs with different $n$ at the initial plane, plotted in logarithmic scales. Other parameters are same as those in Fig. S8.

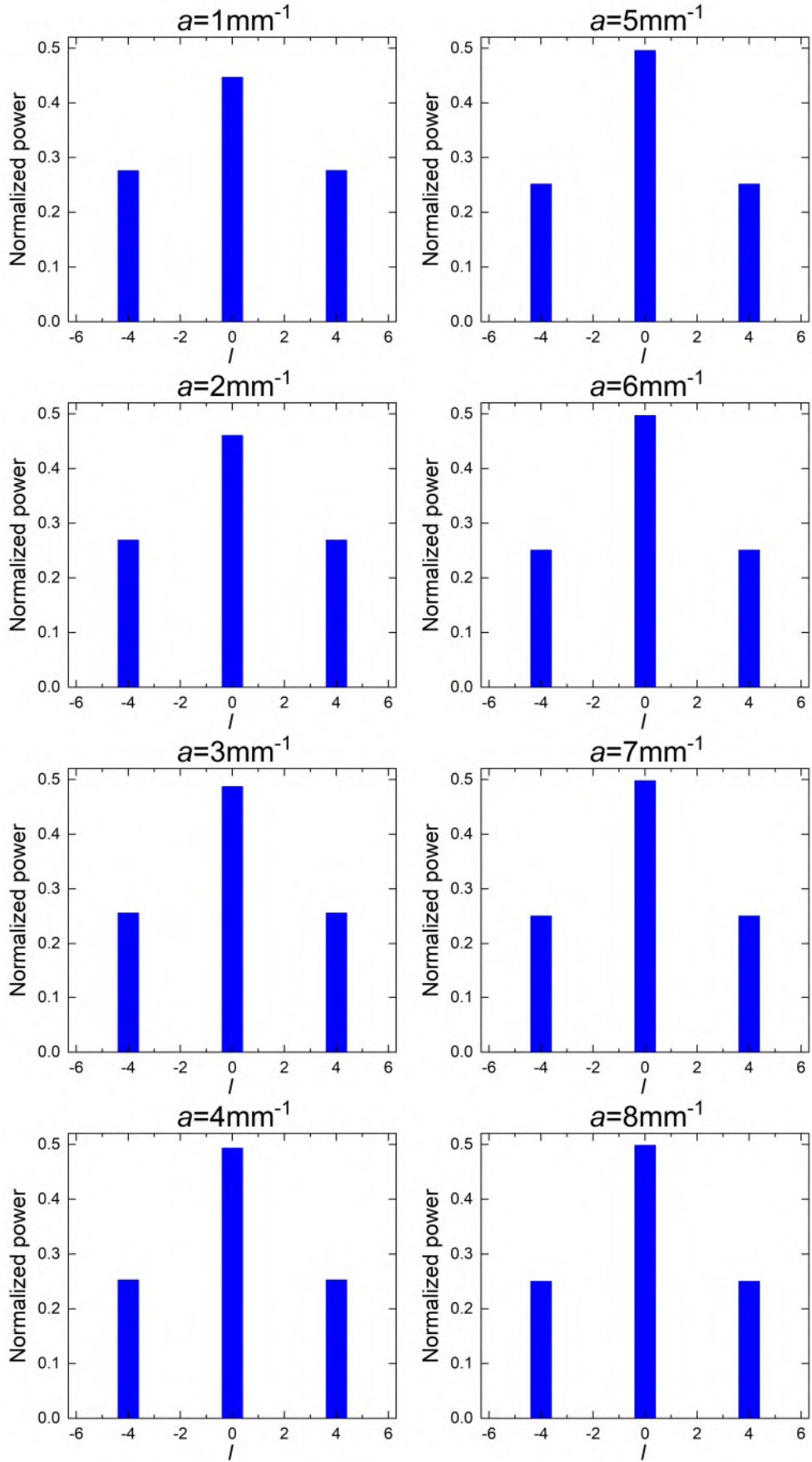

**Fig. S10.** OAM spectra of the CSAOVLs with $n = 1$ at the initial plane, under different $a$. Other parameters are $b = 4$, and $w_0 = 1.63$ mm.

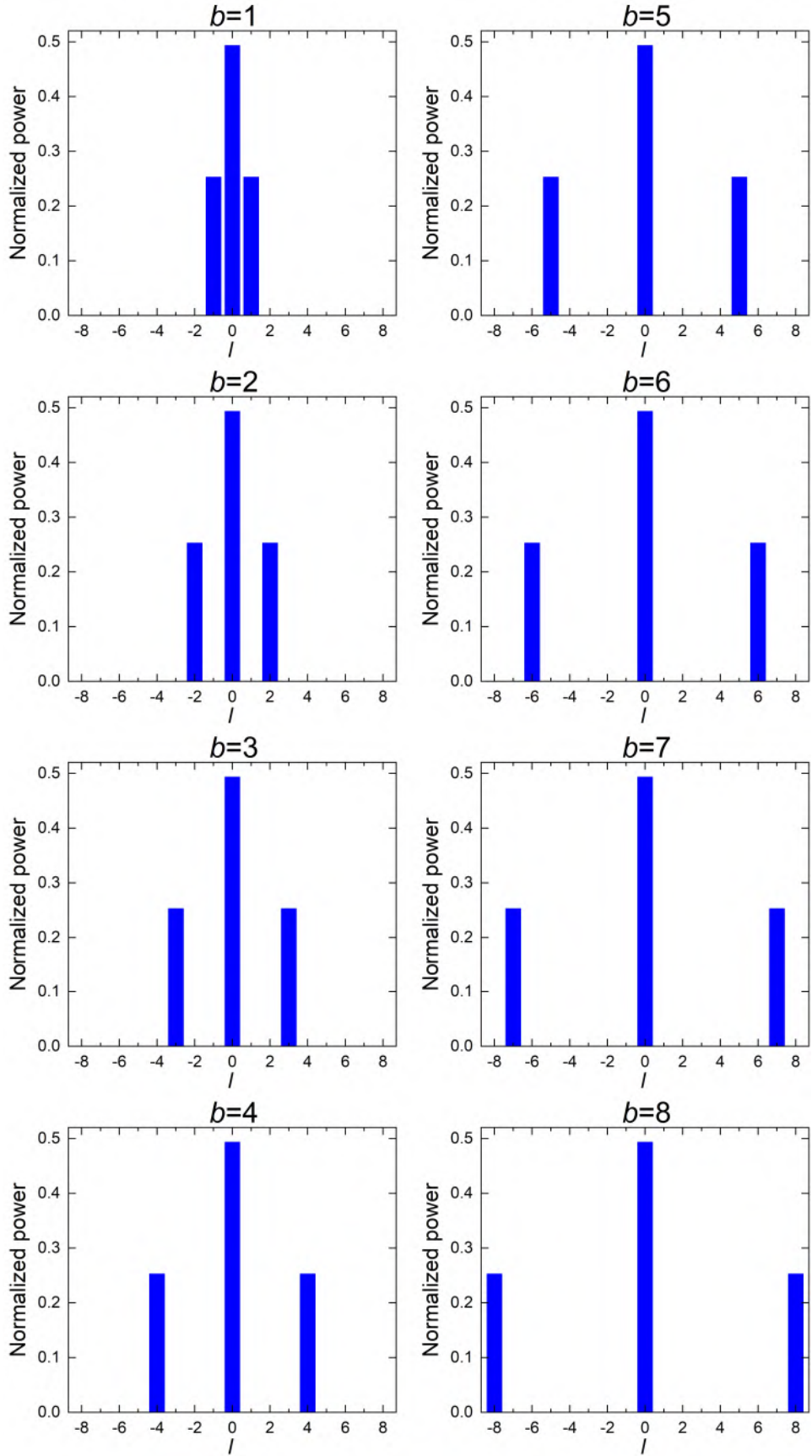

**Fig. S11.** OAM spectra of the CSAOVLs with $n = 1$ at the initial plane, under different $b$. Other parameters are $a = 4$ mm$^{-1}$, and $w_0 =$ 1.63 mm.

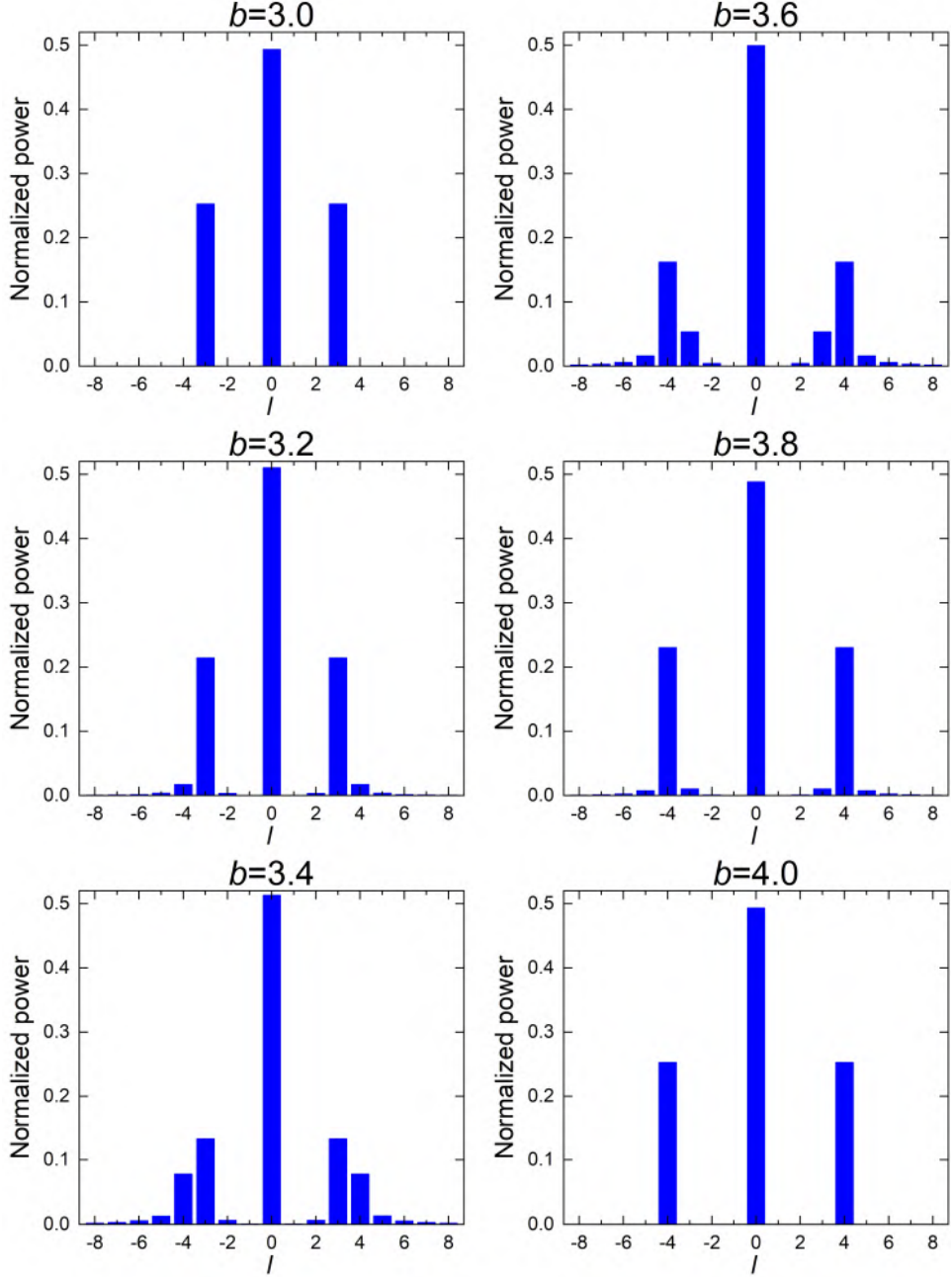

**Fig. S12.** OAM spectra of the CSAOVLs with $n = 1$ at the initial plane, for $b$ changing from 3 to 4. Other parameters are $a = 4$ mm$^{-1}$, and $w_0 = 1.63$ mm.

## E. Additional intensity distributions of the CSAOVLs with fractional values of $b$ at the focal plane of a 2-$f$ lens system, under different $n$

In Figs. 2-4, it is observed that when $b$ is a positive integer, the intensity patterns of the circular optical arrays exhibit diverse rotational symmetries. To examine the effects of fractional $b$ on the intensity patterns, Fig. S13 is presented to illustrate the intensity distributions of circular optical arrays with different fractional values of $b$. When the value of $b$ deviates slightly from an integer value, as depicted in Figs. S13(a1)-S13(d1) or Figs. S13(a5)-S13(d5), the distortions in the circular optical arrays are subtle. With an increasing degree of deviation of $b$ from an integer value, the distortion phenomenon in the circular optical arrays becomes more pronounced, as shown in Figs. S13(a2)-S13(d2) or Figs. S13(a4)-S13(d4). When $b$ is a half-integer, distinct distortions of bright spots in the circular optical arrays are evident, as presented in Figs. S13(a3)-S13(d3). These distortions in the bright spots of circular optical arrays are non-rotationally symmetrical, contributing to the formation of more intricate and non-rotationally symmetrical intensity patterns. Consequently, a fractional value of $b$ can introduce more complex intensity structures.

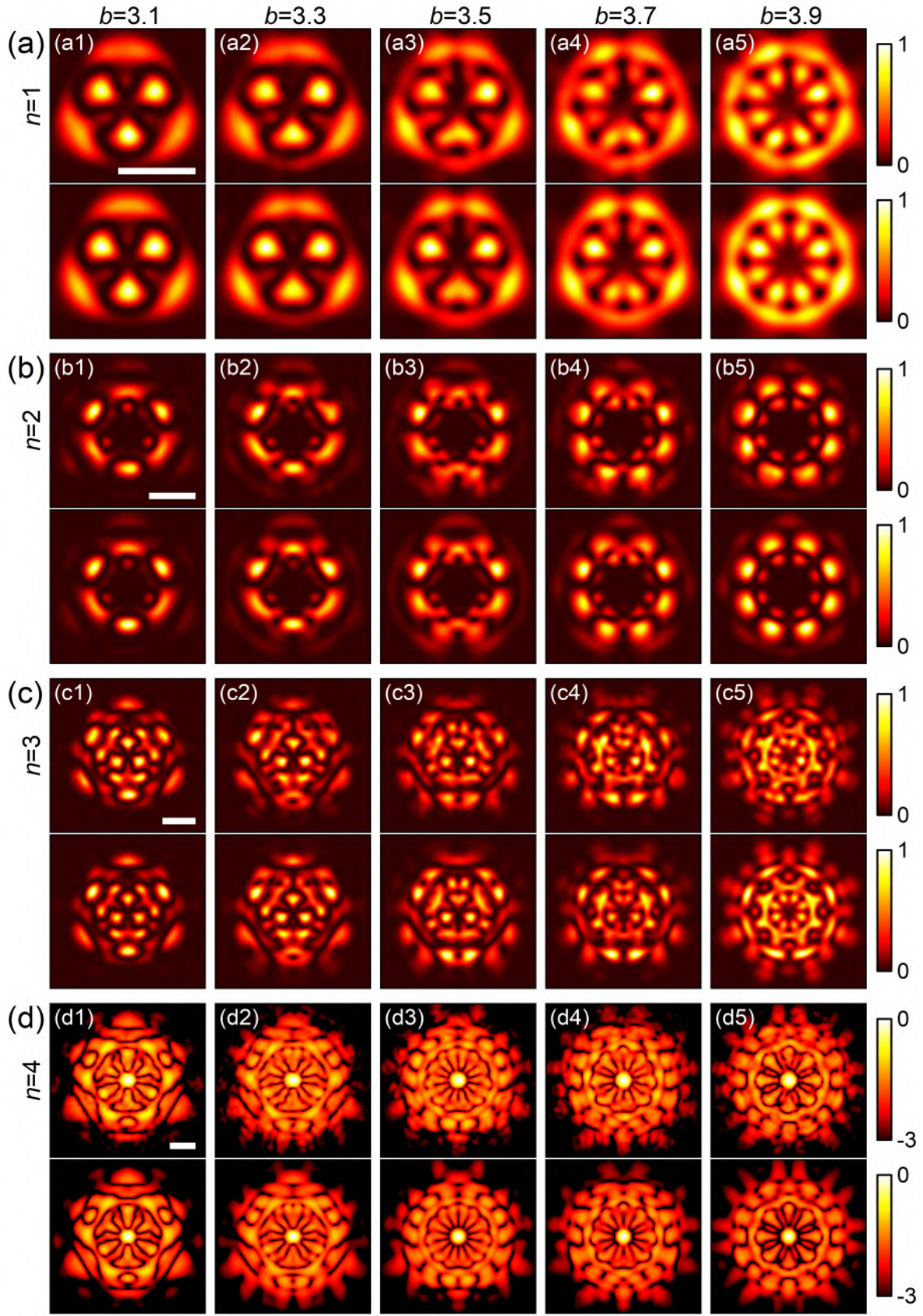

**Fig. S13.** Experimental (upper) and theoretical (lower) intensity distributions with different fractional $b$ at the focal plane of a 2-$f$ lens system, for four different CSAOVLs with (a) $n = 1$, (b) $n = 2$, (c) $n = 3$, and (d) $n = 4$. The intensities in Fig. (d) are plotted in logarithmic scales. The values of $b$ are labeled on the top of subfigures. Other parameters are $a = 4$ mm$^{-1}$, $w_0 = 1.63$ mm, and $f = 500$ mm. White scale bars, 0.3 mm.

# References

[1] S.A. Collins, Lens-system diffraction integral written in terms of matrix optics, J. Opt. Soc. Am. 60 (9) (1970) 1168 –1177.

[2] S. Wang, D. Zhao, Matrix optics, CHEP-Springer (2000).

[3] S. Wen, L.G. Wang, X.H. Yang, J.X. Zhang, S.Y. Zhu, Vortex strength and beam propagation factor of fractional vortex beams, Opt. Express 27 (4) (2019) 5893–5904.

[4] Y.-D. Liu, C. Gao, M. Gao, F. Li, Coherent-mode representation and orbital angular momentum spectrum of partially coherent beam. Opt. Commun. 281, 1968-1975 (2008).

[5] C. Schulze, A. Dudley, D. Flamm, M. Duparre, A. Forbes, Measurement of the orbital angular momentum density of light by modal decomposition. New J. Phys. 15, 073025 (2013).

[6] L. Torner, J.P. Torres, S. Carrasco, Digital spiral imaging. Opt. Express 13, 873-881 (2005).

[7] N. Matsumoto, T. Ando, T. Inoue, Y. Ohtake, N. Fukuchi, T. Hara, Generation of high-quality higher-order Laguerre-Gaussian beams using liquid-crystal-on-silicon spatial light modulators. J. Opt. Soc. Am. A 25, 1642-1651 (2008).